%% file: main.tex
\documentclass[reprint,superscriptaddress,amsmath,amssymb,aps]{revtex4-2}
\usepackage{graphicx}% Include figure files
\usepackage{float}
\usepackage{dcolumn}% Align table columns on decimal point
\usepackage{bm}% bold math
\usepackage{physics}
\usepackage{xcolor}
\definecolor{darkblue}{rgb}{0,0,0.6}
\usepackage[colorlinks,linkcolor=darkblue,citecolor=darkblue,urlcolor=darkblue]{hyperref}

\newcommand{\montpellier}{Laboratoire Charles Coulomb (L2C), Universit\'e de Montpellier, CNRS, 34095 Montpellier, France}
\newcommand{\gulliver}{Gulliver, UMR CNRS 7083, ESPCI Paris, PSL Research University, 75005 Paris, France}

\usepackage[normalem]{ulem}

\begin{document}

\title{Direct numerical analysis of dynamic facilitation in glass-forming liquids}

\author{Cecilia Herrero}

\affiliation{\montpellier}

\author{Ludovic Berthier}

\affiliation{\montpellier}
\affiliation{\gulliver}

\date{\today}

\begin{abstract}
We propose a computational strategy to quantify the temperature evolution of the timescales and lengthscales over which dynamic facilitation affects the relaxation dynamics of glass-forming liquids at low temperatures, that requires no assumption about the nature of the dynamics. In two glass models, we find that dynamic facilitation depends strongly on temperature, leading to a subdiffusive spreading of relaxation events which we characterize using a temperature-dependent dynamic exponent. We also establish that this temperature evolution represents a major contribution to the increase of the structural relaxation time. 
\end{abstract}

\maketitle

%{\it broad intro to dynamics}
The relaxation dynamics of glass-formers in the vicinity of the experimental glass transition is not fully elucidated~\cite{ediger1996,ediger2000spatially,debenedetti2001supercooled}. It is often difficult to draw firm conclusions about the relative role of distinct mechanisms from measurements without relying on unproven hypothesis~\cite{berthier2011}. Our main goal is to quantitatively assess the role and temperature evolution of dynamic facilitation~\cite{palmer1984,fredrickson1984kinetic,jackle1991,chandler2010dynamics} in the relaxation of glass-forming liquids without making assumptions about the nature of relaxation events.  

%{\it dynamic facilitation is invoked by theories}
Dynamic facilitation captures the physical idea that a relaxation event happening somewhere causally triggers future relaxation events. It is invoked in several theoretical approaches~\cite{xia2001microscopic,mirigian2013unified,schoenholz2016structural}. Some models and theories are directly constructed around facilitation~\cite{fredrickson1984kinetic,jackle1991,chandler2010dynamics,speck2019}, thought to be triggered by localised mobility defects. It has been suggested that elasticity may be responsible for mediating dynamic information~\cite{mirigian2013unified,ozawa2023,tahaei2023,ghanekarade2023}. However, there exists no first-principles description of dynamic facilitation to predict its strength and temperature evolution for atomistic models.  

%{\it Facilitation was observed before}
Progress will require quantitative observations from realistic models. Previous work suggested the existence of dynamic facilitation by detecting cross-correlations between successive relaxation events~\cite{vogel2004spatially,bergroth2005examination,keys2011,gokhale2014growing,chacko2021}, but this approach remains qualitative. In \cite{keys2011,isobe2016,hasyim2021theory}, the relaxation dynamics was deemed hierarchical, which amounts to the logarithmic growth of energy barriers with distance. Invoking the Arrhenius law, this is mathematically equivalent to a power-law relation between timescales and lengthscales, $t \sim \ell^{z}$, with a temperature-dependent dynamic exponent, $z(T) \sim 1/T$, as explicitly found in certain kinetically constrained models~\cite{jackle1991,sollich1999glassy,berthier2005numerical}. 

The large body of data accumulated in studies of four-point correlations~\cite{donati1999,lacevic2003,toninelli2005,karmakar2010} for dynamic heterogeneity~\cite{berthier2011dynamical} can potentially elucidate the relation between space and time~\cite{berthier2005numerical}. However, their interpretation is not unique~\cite{toninelli2005}, as multi-point functions do not directly probe the underlying relaxation mechanisms. At times shorter than the structural relaxation, this problem was circumvented by introducing a growing lengthscale characterizing the subdiffusive spreading of mobile regions~\cite{scalliet2022}.

The emergence of propagating fronts in ultrastable glasses heterogeneously transforming into supercooled liquids~\cite{ediger2017highly, herrero2023front} is a form of dynamic facilitation~\cite{leonard2010,sepulveda2013manipulating}, but the non-equilibrium nature of the relaxation leads to a ballistic ($\ell \propto t$) growth of relaxed regions~\cite{herrero2023two}, unlike the subdiffusion found in equilibrium. 

\begin{figure}
 \includegraphics[width=\linewidth]{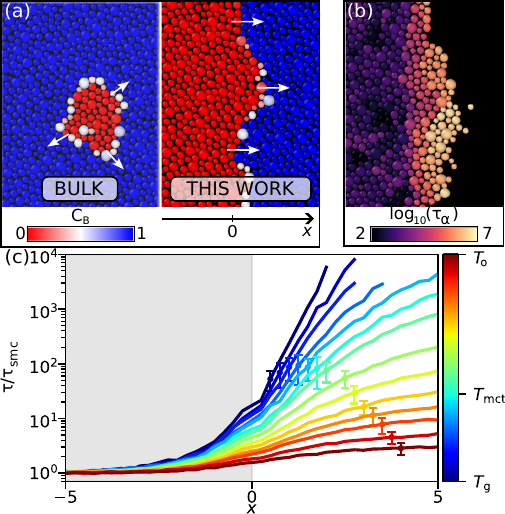}
 \caption{{\bf Direct measurement of dynamic facilitation.}
   (a) Instead of measuring the growth of rare mobile regions in the bulk (left), we create a macroscopic interface (right) separating a mobile region for $x<0$ (using swap MC) from a region at $x>0$ where conventional MD at the same $T$ is used. The system is structurally homogeneous at all times. 
   (b) Map of local relaxation times near $x=0$ showing the spatial spreading of mobility along the horizontal axis ($T=0.09$). 
   (c) Dynamic profiles showing the evolution of the structural relaxation with $x$, normalised by $\tau_{\rm smc}$, for the $2d$ system. Dynamic facilitation becomes less efficient at lower $T$. Typical errorbars are shown at some selected data point for each temperature. Full errorbars are shown in the SM~\cite{supplement}}
 \label{fig:fig1}
\end{figure}

%{\it our strategy}
Here we propose a computational strategy to understand if and how much dynamic facilitation affects the equilibrium dynamics of deeply supercooled liquids. We conduct simulations where a macroscopic interface separates two regions evolving with distinct equilibrium dynamics at the same temperature $T$: ordinary molecular dynamics (MD) in one domain, and swap Monte Carlo (SMC) dynamics in the other. The rationale is that in bulk dynamics (Fig.~\ref{fig:fig1}a), rare mobile regions appear whose spatial spreading reveals facilitation~\cite{scalliet2022}. Here instead, SMC dynamics creates a macroscopic mobile region whose influence then spreads to the MD region by facilitation in a controlled geometry (Fig.~\ref{fig:fig1}a). Since both regions are equilibrated at the same $T$, they are structurally indistinguishable at any time and the system is stationary. Therefore, the fast dynamics in the SMC region facilitates relaxation in the MD region. The observed relaxation is representative of the equilibrium dynamics, but the measurement of a macroscopic mobility front (Fig.~\ref{fig:fig1}b) does not rely on any assumption about the nature of the microscopic events or on detailed knowledge of the origin of facilitation. This allows us to directly access the relation between space and time over a broad range of temperatures (Fig.~\ref{fig:fig1}c).

%{\it methods, models details}
We perform simulations in 2 and 3 dimensions of soft size-polydisperse spheres~\cite{berthier2017configurational,berthier2019zero}, with the same size polydispersity $\delta = 23\%$ (see SM~\cite{supplement}).
%. 
The pairwise interaction potential is:
\begin{equation}
    V_{ij}(r_{ij}) = \varepsilon \qty( \frac{\sigma_{ij}}{r_{ij}} )^{12} + c_0 + c_2 \qty( \frac{r_{ij}}{
    \sigma_{ij}} )^2 + c_4 \qty( \frac{r_{ij}}{\sigma_{ij}} )^4,
\end{equation}
where $r_{ij}$ is the interparticle distance and $\sigma_{ij} = \frac{\sigma_i + \sigma_j}{2}(1 - \eta \abs{\sigma_i - \sigma_j})$, with $\eta=0.2$ a non-additivity parameter. The parameters $c_0 = -28 \varepsilon/r_{\rm c}^{12}$, $c_2 = 48 \varepsilon/r_{\rm c}^{14}$, and $c_4 = - 21 \varepsilon/r_{\rm c}^{16}$, ensure the continuity of the potential up to its second derivative at the cut-off distance $r_{\rm c} = 1.25\sigma_{ij}$. We use reduced units based on the particle mass $m$, the energy scale $\varepsilon$, and the average particle diameter $\sigma$, so the time unit is $\tau =  \sigma \sqrt{m/\varepsilon}$. 

We first equilibrate the entire system at the desired temperature $T$ and number density $\rho = 1$ using swap Monte Carlo (SMC)~\cite{ninarello2017,berthier2019efficient,berthier2019zero}. We then impose two distinct dynamics in two regions of space, keeping the periodic boundary conditions. For $x<0$ we run SMC using the hybrid scheme developed in \cite{berthier2019efficient}. For $x>0$, we run standard molecular dynamics (MD). Due to the periodic boundary conditions the mobile SMC region forms a slab. As rationalised below, we found it convenient to impose SMC on 25\% of the system, and MD in the rest. Simulations are performed at constant $T$ using a Nos\'e-Hoover thermostat with a timestep of 0.01. We also varied the amount of swap MC moves~\cite{berthier2019efficient} for $x<0$ to assess its role in the quantification of dynamic facilitation. Additional details and several tests are provided in SM~\cite{supplement}.

We adopt the usual definitions of characteristic temperature scales~\cite{scalliet2022}. In $3d$, the onset temperature is $T_o=0.2$, the mode coupling crossover $T_{\rm mct} = 0.095$, and the glass transition temperature $T_g = 0.056$. The total number of particles is $N=96000$ and we explored temperatures in the range $T\in [0.075,0.2] $. In $2d$, $T_o=0.2$, $T_{\rm mct} = 0.12$, $T_g = 0.07$, and we explored temperatures $T \in [0.07, 0.2]$. The number of particles for $T \ge 0.1$ is $N=10000$. For lower $T$, the relaxation timescale in the MD region is so large that mobility far from the interface is negligible. For $T<0.1$, we thus simulated a smaller system with $N=2500$ reducing only the length of the MD region. When possible, we checked that the two geometries provide similar results. We run simulations up to $t=2 \times 10^7$ (about 2 weeks of CPU time), and perform up to 50 independent simulations per temperature.

We use the bond-breaking correlation to quantify the dynamics:
\begin{equation}
    C_B^i(t) = \frac{n_i(t|0)}{n_i(0)},
    \label{eq:cbi}
\end{equation}
with $n_i(t)$ the number of neighbors of particle $i$ at time $t$, and $n_i(t|0)$ the number of those initial neighbors that remain at time $t$.
We follow \cite{guiselin2022} for the neighbor definitions. At $t=0$, the neighbors of particle $i$ are particles $j$ which are closer than a threshold, $r_{ij} / \sigma_{ij}<1.35$ in $2d$ and $r_{ij}/\sigma_{ij} < 1.49$ in $3d$, which correspond to the first minima of the rescaled radial distribution function $g(r/\sigma_{ij})$. We spatially resolve the dynamics by measuring $C_B(x,t)$ for all particles in the interval $x \pm \delta x/2$, with $\delta x = 0.25$. We define the relaxation time $\tau(x,T)$ as  $C_{B}(x,\tau) = 0.7$. We obtained equivalent results for different thresholds. We denote $\tau_{\rm smc}$ and $\tau_\alpha$ the bulk relaxation times for SMC and MD dynamics, respectively. 

%{\it first, we observe facilitation}
Dynamic facilitation is demonstrated in Fig.~\ref{fig:fig1}b, where a gradient of relaxation times is observed near the macroscopic interface at $x=0$, showing that the mobile regions at $x<0$ indeed trigger the relaxation of particles following the conventional MD dynamics at $x>0$. Because the structure is completely homogeneous along $x$ at all times (see SM~\cite{supplement}), this acceleration necessarily results from dynamic facilitation. Dynamic profiles $\tau(x,T)$ are reported in Fig.~\ref{fig:fig1}c, using $\tau_{\rm smc}$ as normalisation. These data reveal that the spatial spreading of mobility becomes less efficient, and thus much slower, at lower temperature. See SM~\cite{supplement} for an equivalent observation in $3d$. 

%{\it then we measure the relation between time and length}
Normalising the data by $\tau_{\rm smc}$ simply shifts the data vertically by an amount $\tau_{\rm smc}$ which depends on the details of the swap Monte Carlo simulations. We have run simulations changing the parameters of the SMC dynamics, and confirmed that the normalised profiles $\tau(x,T)$ are unchanged, see SM~\cite{supplement}). When $T$ is not too low we can observe plateaus in both regions when $|x| \gg 1$. In particular, convergence to the MD plateau occurs for $x \sim \xi_d$, where $\xi_d(T)$ is the dynamic correlation length~\cite{kob2012}. At low temperatures, $\tau_\alpha$ becomes too large and we cannot follow the profiles up to $\tau_\alpha$ in the MD region. Crucially, however, since relaxation remains fast in the SMC region, we can still characterise dynamic facilitation over several decades in time down to $T_g$.

\begin{figure}
 \includegraphics[width=\linewidth]{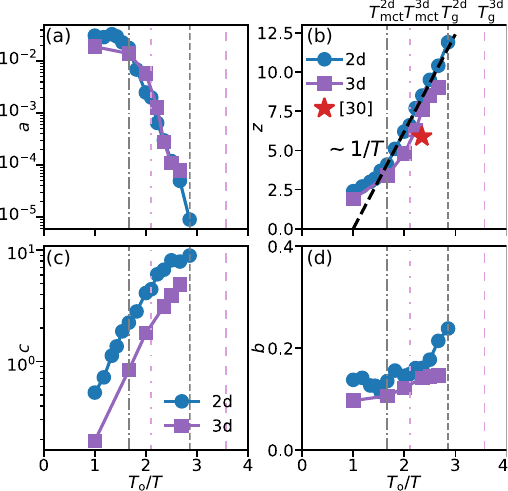}
 \caption{{\bf Temperature evolution of space-time relation.} Evolution of fitting parameters in $2d$ and $3d$ models. 
   (a, b) Parameters for power law fit. The red star in (b) corresponds to the measurement in Ref.~\cite{scalliet2022}, and the black dashed line to a $z \sim 1/T$ behaviour.
   (c, d) Parameters for logarithmic growth.}
 \label{fig:fig2}
\end{figure}

%{\it we quantify our findings for intermediate distances}
To quantify these observations, we describe for $0 < x < \xi_d$ subdiffusive spreading of mobility from fast to slow regions using either a power law, 
\begin{equation}
x \sim \tau(x,T)^{1/z(T)},
\label{eq:z}
\end{equation}
where $z(T)$ is a temperature dependent dynamic exponent, or a thermally activated logarithmic form:
\begin{equation}
    x \sim T \log \tau(x,T).
    \label{eq:log}
\end{equation}
For $x \approx \xi_d$, we expect a saturation to $\tau_\alpha$, which we capture using
\begin{equation}
    \tau(x,T) \sim \tau_\alpha (1 - e^{- x / \xi_d}).
    \label{eq:length_fit}
\end{equation}

In practice, we account for the crossover to $\tau_{\rm smc}$ at $x<0$ using the expression $\tau(x) / \tau_{\rm smc} = 1 + a(T) ( x + x_0 )^{z(T)}$ where $a$, $x_0$ and $z(T)$ are fitting parameters. This reduces to Eq.~(\ref{eq:z}) in the relevant regime $\tau_{\rm smc} \ll \tau(x,T) \ll \tau_\alpha$. We found that fixing $x_0 = 3.3$ in $2d$ and $x_0 = 3.4$ in $3d$ described the data well. The quality of the fits can be appreciated in SM~\cite{supplement}. The temperature evolution of $a$ and $z$ are shown in Fig.~\ref{fig:fig2}(a,b) using the rescaled axis $T_o/T$ to compare both $2d$ and $3d$ models. We first notice the very comparable evolution obtained for the two models, which reveals the weak influence of the spatial dimension. The temperature evolution of $a$ mostly mirrors the one of $\tau_{\rm smc}$, as it should. The most interesting observation is the evolution of $z(T)$ which reveals nearly diffusive behaviour ($z \sim 2$) near $T_o$, but increases very fast as $T$ is lowered reaching the large value $z \approx 12$ near $T_g$ in $2d$. The rapid growth of $z(T)$ captures the evolution of dynamic facilitation towards much slower mobility propagation at low temperatures. Our data are compatible with $z(T) \approx A/T$ at low $T$, with $A$ some constant. The agreement with the single data point reported in Ref.~\cite{scalliet2022}, determined in the bulk geometry for the same $2d$ model, is quite good, given the difference in methodologies. We also characterised the growth of isolated domains directly in the bulk (see SM~\cite{supplement}), and found good agreement with the $z(T)$ results reported in Fig.~\ref{fig:fig2}b. This provides quantitative support to the analogy between the geometries illustrated in Fig.~\ref{fig:fig1}a.

Given the large values of $z$, it is natural that a logarithmic growth can also be used, although the linear behaviour predicted in Eq.~(\ref{eq:log}) is not obvious in the profiles shown in Fig.~\ref{fig:fig1}c. In practice, we fit the data to the functional form $\tau(x,T) / \tau_{\rm smc} = 1 + c(T) \exp(b(T) x/T)$ with $c$ and $b$ fitting parameters. This expression reduces to Eq.~(\ref{eq:log}) far from the plateaus. This second form fits a slightly smaller $x$-range but describes the evolution of the dynamic profiles reasonably well, as shown in SM~\cite{supplement}. We report the mild, but non-negligible, evolution of the parameters $b$ and $c$ in Fig.~\ref{fig:fig2}(c,d) which again reveal a similar evolution for both models. It is not surprising that both power-law and logarithmic functional forms can fit the data at low temperatures, as they are mathematically close when $z(T) \approx A/T$ since $x \sim \tau^{1/z(T)} = e^{ (T/A) \log\tau } \sim 1 + (T/A) \log \tau$. This coincidence was noted before~\cite{berthier2002geometrical,dalle2007spatial}.   

\begin{figure*}
\includegraphics[width=\linewidth]{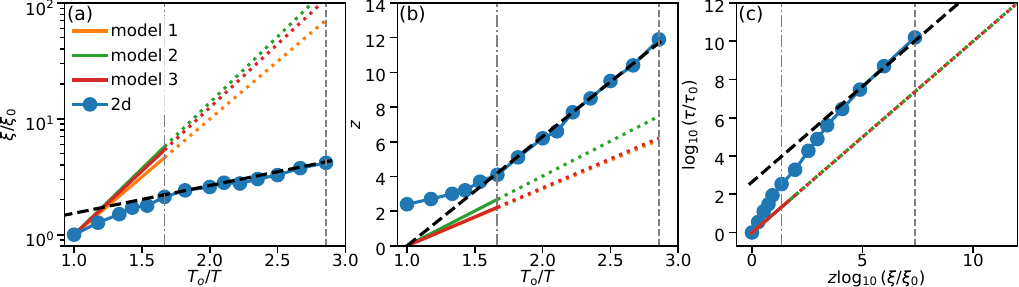}
\caption{{\bf Understanding the temperature evolution of structural relaxation.} Simulation results for the 2d system (circles), compared to results for three different models from Ref.~\cite{keys2011} (full lines are actual results, dotted lines are continuation to $T_g$ using asymptotic expressions). (a) Dynamic correlation length $\xi_d$ normalized by its onset value as a function of $T_o/T$. Black dashed line is an Arrhenius fit. (b) Dynamic exponent $z$ as a function of $T_o/T$. Black dashed line is $z(T) \sim A/T$. (c) Parametric plot of $\tau_\alpha$ against its facilitation estimate $\xi_d^z$. Black dashed line indicate linear relation, which leaves an offset of about 2 decades.}  
\label{fig:fig3}
\end{figure*}

% we extract the characteristic dynamic lengthscale
We finally use Eq.~(\ref{eq:length_fit}) to extract the dynamic correlation lengthscale $\xi_d$. At very low temperatures where the MD plateau cannot be reached, we use an extrapolation of the bulk relaxation times $\tau_\alpha(T)$ using a parabolic law~\cite{berthier2020}. We then estimate $\xi_d$ from the fit to the dynamic profile, assuming $\tau(x=\xi_d,T)=\tau_\alpha(T)$. We verified that both approaches (direct measurement and extrapolation) compare well when both can be used. The temperature evolution of $\xi_d$, normalized by its value at the onset temperature $\xi_0=\xi_d(T_o)$, is shown in Fig.~\ref{fig:fig3}a. After a fast transient at high temperature, we observe that the characteristic dynamic lengthscale is well described by an Arrhenius form, $\xi_d/\xi_0 \sim \exp(E_d/T)$, with $E_d \approx 0.13$, leading to an overall increase of a factor 4 between $T_o$ and $T_g$ in $2d$. Our estimate for $\xi_d$ compares very well, but extends to lower $T$, the evolution of the average chord length reported for the same $2d$ model~\cite{scalliet2022}. Given the modest evolution of $\xi_d$, other functional forms are presumably possible.

% {\it comparison to previous work for z(T)}
We now compare our findings with earlier work. The good agreement with the determination of $z(T)$ from the average chord length measured in bulk simulations~\cite{scalliet2022} not only confirms the validity of our strategy, but also shows that our approach is more flexible and much easier to implement over a broader temperature range.

Reference \cite{keys2011} proposes, among other measurements, a strategy to relate the extent of spatial relaxation events to their timescale. We can recast these results into a dynamic exponent, $z(T)= \alpha (1/T - 1/T_o)$ with a prefactor $\alpha$ that was evaluated for several models~\cite{keys2011}. We explain this dictionary in SM~\cite{supplement}. We compare these results to ours in Fig.~\ref{fig:fig3}b. The two data sets deviate on two aspects. First, our data do not indicate that $z$ vanishes at $T_o$. Second, our approach yields considerably larger values at low $T$: both sets yield $z \approx A/T$ at low $T$ but with different prefactors. Measurements along the lines of \cite{keys2011} should be performed at lower $T$ to better assess the nature of the reported difference in this regime.

%{\it then we compare our length to their concentration}
A second comparison point with \cite{keys2011} is the characteristic length $\xi_d$. While we directly measured it, Keys {\it et al.} assume instead that relaxation events are triggered by localised excitations and estimate the average distance between them. We report tabulated values~\cite{keys2011} in Fig.~\ref{fig:fig3}a. While both data sets suggest a similar Arrhenius dependence, the corresponding energy scales $E_d$ differ considerably. The very large correlation length predicted by Keys {\it et al.} at low $T$ appears unrealistic~\cite{dalle2007spatial}. We can conclude that either the numerical technique used before~\cite{keys2011,isobe2016,gokhale2014growing,hasyim2021theory,ortlieb2023} to estimate the concentration of excitations is too crude, or that relaxation events are not simply due to localised excitations~\cite{scalliet2022}. The very low $T$ measurements in \cite{ortlieb2023} do not show sign of a temperature crossover which could reconcile both families of measurements at low $T$.

%{\it we test whether a picture of slow growth of domains of size $\xid$ explains all of $\tau_\alpha$}
Our results allow us to test the validity of the facilitation picture. Assuming that localised excitations relaxing with a characteristic timescale $\tau_{\rm ex}(T) \ll \tau_\alpha(T)$ can relax the entire system via dynamic facilitation, we arrive at
\begin{equation}
\tau_\alpha(T) \sim \tau_{\rm ex} \xi_d^{z},
\label{eq:prediction}
\end{equation}
which represents the time it takes to grow a domain of size $\xi_d$ from a localised excitation via dynamic facilitation. We test Eq.~(\ref{eq:prediction}) using our direct, agnostic determinations of $z$, $\xi_d$, and $\tau_\alpha$. The results in Fig.~\ref{fig:fig3}c show that after a short transient at high $T$, $\tau_\alpha$ becomes indeed proportional to $\xi_d^{z}$, with a prefactor of about $10^2$ near $T_g$. This result suggests that the joint temperature evolution of $\xi_d(T)$ and $z(T)$ accounts for most of the 12-decade slowdown of the structural relaxation time $\tau_\alpha(T)$. 

This conclusion is significant because it demonstrates the central role played by dynamic facilitation in controlling the dynamic slowdown of deeply supercooled liquids. At any temperature, dynamic facilitation `accelerates' the relaxation via the successive triggering of relaxation events~\cite{guiselin2022}, but since this process becomes increasingly inefficient at low $T$ (as captured by the rapid growth of $z$) the overall relaxation takes a longer time.

%{\it link with fragility}
Our study thus identifies the two major contributors, $z(T)$ and $\xi_d(T)$, to the temperature dependence of $\tau_\alpha(T)$. While the interpretation of $z$ is clear (it quantifies facilitation), our approach does not explain the evolution of $\xi_d(T)$ shown in Fig.~\ref{fig:fig3}a, and this should be the subject of future studies. Combining the asymptotic Arrhenius evolution of $\xi_d$ to the $1/T$ dependence of $z$ found numerically provides an expression for the relaxation time: 
\begin{equation}
    \tau_\alpha \sim \xi_d^z \sim \exp \left( \frac{E_d A}{T^2} \right),
    \label{eq:bassler}
\end{equation}
which is the parabolic B\"assler law~\cite{bassler1987viscous}. While \cite{keys2011} arrived at a similar functional form, we noted above that our determinations of $E_d$ and $A$ differ quantitatively, even if the products $E_d A$ in Eq.~(\ref{eq:bassler}) appear very close to those reported in Ref.~\cite{keys2011}, as seen in Fig.~\ref{fig:fig3}(c).

%{\it The DW factor}
The proposed slab geometry allows us to quantify how fast mobile regions spread to immobile ones with no assumption about the underlying microscopic mechanism. We can however test the specific modelling of facilitation of \cite{mirigian2013unified} which assumes that local elasticity is responsible for dynamic facilitation. As noted recently in thin polymer films~\cite{ghanekarade2023}, the slab geometry yields an algebraic gradient of elastic moduli, thus leading to algebraic dynamic profiles. We have tested this model by measuring spatial profiles of the Debye-Waller factor (mean-squared displacement at short times, known to strongly correlate with the shear modulus), and found rapid (non-algebraic) convergence to the bulk value (see SM~\cite{supplement}), in agreement with earlier results~\cite{sussman2017}. There is thus no correlation between elasticity and dynamic profiles, which are also found to converge exponentially (not algebraically) to the bulk behaviour [recall Eq.~(\ref{eq:length_fit})]. Together with \cite{sussman2017}, our data do not follow the analytic description of Ref.~\cite{mirigian2013unified}. It would be interesting to analyse thermal elasto-plastic models~\cite{ozawa2023} in slab geometries. Although more limited, our results also suggest that the rugosity of the dynamic profiles decreases at low temperatures (see SM~\cite{supplement}). This is in agreement with earlier numerical findings~\cite{scalliet2022,DAS2022100098,jung2023predicting}, but differs from hierarchical kinetically constrained models~\cite{berthier2005numerical} and geometries found in elasto-plastic models~\cite{ozawa2023}. 
 
%{\it it's time to conclude}
In conclusion, we introduced a computational scheme to quantify the role of dynamic facilitation in the relaxation dynamics of deeply supercooled liquids, which demonstrates a strongly subdiffusive and temperature dependent spatial spreading of relaxation events. A first future task is to expand these studies to a broader range of models to validate the generality of our findings. Our results also suggest that extending the approach of \cite{keys2011} to lower temperatures would be instructive. Future work should elucidate the microscopic mechanisms giving rise to the particular temperature dependence of the dynamic exponent $z(T)$. Finally, it is unclear how these results can be reconciled with the idea that dynamics proceeds via cooperative activated events stemming from static configurational fluctuations~\cite{kirkpatrick1989,bouchaud2004,berthier2021}.
% A possible hypothesis would be that static fluctuations play a role in controlling the temperature evolution of the dynamic correlation lengthscale $\xi_d$.
Overall, our study considerably sharpens the set of questions to be addressed in order to fully elucidate the nature of the structural relaxation in supercooled liquids near the glass transition.

\acknowledgments

We thank G. Biroli, J.-P. Bouchaud, J. Kurchan, and G. Tarjus for discussions. This work was funded through ANR (the French National Research Agency) under the Investissements d'avenir programme with the reference ANR-16-IDEX-0006. It was also supported by a grant from the Simons Foundation (\#454933, LB), and by a Visiting Professorship from the Leverhulme Trust (VP1-2019-029, LB). 

%\bibliography{cecilia.bib}
\input{main.bbl}
\end{document}

\title{Supplemental Material: Direct numerical analysis of dynamic facilitation in glass-forming liquids}
	
	\author{Cecilia Herrero}
	
	\affiliation{\montpellier}
	
	\author{Ludovic Berthier}
	
	\affiliation{\montpellier}
	\affiliation{\gulliver}
	
	\date{\today}

	\maketitle
	
	%%%%%%%%%% Merge with supplemental materials %%%%%%%%%%
	%%%%%%%%%% Prefix a "S" to all equations, figures, tables and reset the counter %%%%%%%%%%
	\setcounter{equation}{0}
	\setcounter{figure}{0}
	\setcounter{table}{0}
	\setcounter{page}{1}
	%\makeatletter
	\renewcommand{\theequation}{S\arabic{equation}}
	\renewcommand{\thefigure}{S\arabic{figure}}
	\renewcommand{\bibnumfmt}[1]{[S#1]}
	\renewcommand{\citenumfont}[1]{S#1}
	\renewcommand\thesubsection{\arabic{subsection}}
	\renewcommand\thesubsubsection{\arabic{subsection}.\arabic{subsubsection}  }
	
	%%%%%%%%%% Prefix a "S" to all equations, figures, tables and reset the counter %%%%%%%%%%

	\textbf{Size-polydisperse model glass-former.}
	
	We conduct molecular dynamics simulations on a size-polydisperse mixture in both two and three dimensions, extensively characterised~\cite{ozawa2019,scalliet2022}. The system comprises soft repulsive spheres with diameters $\sigma_i$, distributed accordingly to the probability distribution $\mathcal{P}(\sigma_i) = A \sigma_i^{-3}$, where $A$ is a normalization constant. The diameters $\sigma_i$ range between $\sigma_\mathrm{min}$ and $ \sigma_\mathrm{max}$, with $\sigma_\mathrm{min}/\sigma_\mathrm{max} = 0.45$ and $\sigma_\mathrm{max}=1.62$. The size polydispersity is quantified from the normalized root-mean-square deviation, as $\delta = \sqrt{\langle \sigma^2 \rangle - \langle \sigma \rangle^2}/\langle \sigma \rangle$, where $\langle \sigma \rangle$ and $\langle \sigma^2 \rangle$ are respectively the first and second momentum of the probability distribution function, defined as $\langle ... \rangle = \int \mathcal{P}(\sigma)(...) \dd \sigma $. The given $\sigma_{\rm min}/\sigma_{\rm max}$ ratio yields $\delta \approx 23\%$. The pair interaction potential is defined as:
	\begin{equation}
		V_{ij} (r) = \varepsilon \qty(\frac{\sigma_{ij}}{r})^{12} + c_0 + c_2 \qty(\frac{r}{\sigma_{ij}})^2 + c_4 \qty(\frac{r}{\sigma_{ij}})^4;
	\end{equation}
	where $r = \abs{\vb{r}_i - \vb{r}_j}$ (with $\vb{r}_i$ denoting the position of particle $i$), and non-additive interactions  $\sigma_{ij}=0.5(\sigma_i + \sigma_j)\qty(1 - 0.2 \abs{\sigma_i - \sigma_j})$. We employ reduced units based on the particle mass $m$, the energy scale $\varepsilon$, and a microscopic length $\sigma$ defined as the average particle diameter. Specifically, the time unit is defined as $\tau_\mathrm{LJ} = \sigma \sqrt{m/\varepsilon}$. To ensure continiuity up to the second derivative at the cutoff distance $r_{\rm c} = 1.25~\sigma_{ij}$, the parameters $c_0 = -28\varepsilon/r_\mathrm{c}^{12}$, $c_2 = 48\varepsilon/r_\mathrm{c}^{14}$, and  $c_4 = -21 \varepsilon/r_\mathrm{c}^{16}$ are introduced.
	
	\hspace{0.5cm}
	
	\textbf{Comparison of the fits of the dynamic profiles.}
	
\begin{figure}[h]
		\centering
		\includegraphics{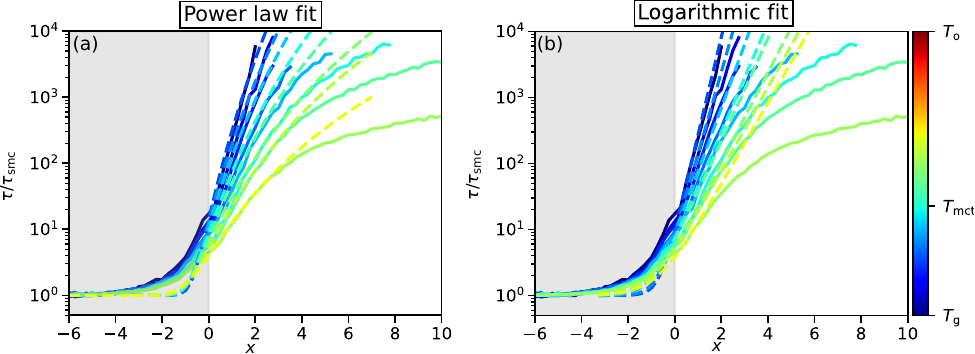}
		\caption{Dynamic profiles for different temperatures (continuous lines), together with the fits (dashed lines). The fit is performed for the atoms closer to the swap-md frontier, located at $x=0$, following a power law fit in (a) and a logarithmic law fit in (b).}
		\label{fig:figS1}
	\end{figure}

	In Fig.~\ref{fig:figS1} we reproduce the dynamics profiles shown in main text together with the two fitting forms used to describe them. This allows to better appreciate the extent and agreement between the data and the fitting functions. We note here that the choice of $x_0 = 3.3$ in the power law fit for the $2d$ simulations serves merely as an effective parameter. Its purpose is to emulate the decrease of mobility within the SMC region near the boundary, and allows to recover well the offset of the MD dynamics at $x=0$, with little influence at $x>0$.
	
	\textbf{Statistical noise and details on the dynamic profiles.}
	
	While our numerical model remains minimal, we provide justification here for the selection of diverse parameter choices within the simulations and analysis. Without care, these parameters can significantly influence the fitting of dynamic profiles, thereby impacting the extraction of power-law and logarithmic fits, as shown in Fig.~\ref{fig:figS1}. 
	
	\begin{figure}[h]
		\centering
		\includegraphics{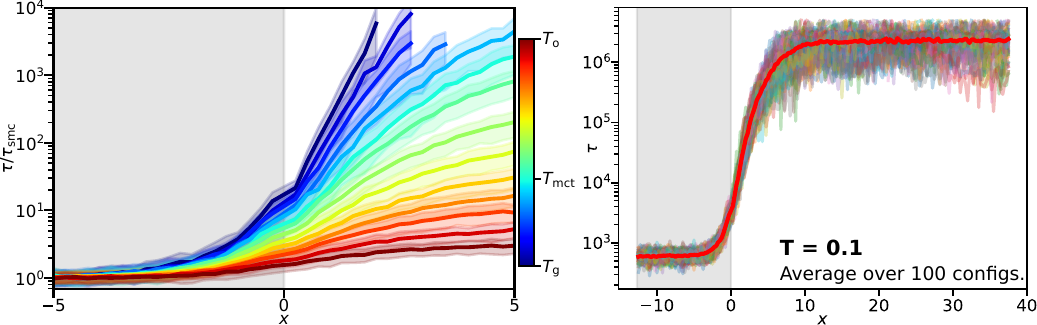}
		\caption{(Left) The dynamic profiles of the $2d$ configuration depicted in Fig.~1 of the main text are presented alongside their corresponding error bars. (Right) Dynamic profile (red) for $T=0.1$ in $2d$ averaged over 100 independent configurations (50 independent runs and 2 sides due to periodic boundary conditions).}
		\label{fig:tx-T01}
	\end{figure}
	
	To reduce the noise in the dynamic profiles, for which the error bars are shown in Fig.~\ref{fig:tx-T01}(left), we averaged over 100 independent configurations. This involved running 50 independent configurations within two SMC-MD interfaces due to periodic boundary conditions. This approach effectively increases considerably the signal to noise ratio, as depicted in Fig.~\ref{fig:tx-T01}(right). 
	
\begin{figure}[h]
		\centering
		\includegraphics{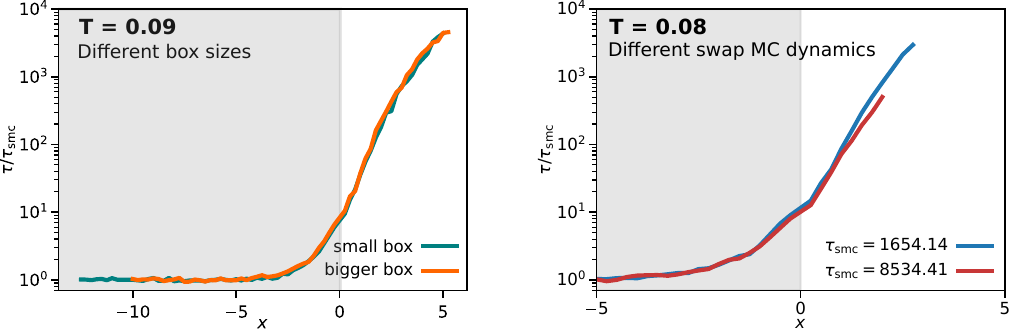}
		\caption{(Left) Robustness of the dynamic profiles measured with different geometries. (Right) Dynamic profile at temperature $T=0.08$ in $2d$ examined under two distinct swap Monte Carlo dynamics. Upon rescaling with the appropriate bulk value within the SMC region, both curves exhibit comparable behavior. This observation highlights the absence of any spurious effects originating from the imposed dynamics in the SMC region on the MD region.}
		\label{fig:rhoswapT008}
	\end{figure}

	Before conducting the fit of dynamic profiles, two important considerations must be addressed: the prescribed dynamics within the SMC region, controlled via the swap density $\rho_{\rm swap}$, and the dimensions of the simulation box. Through the simulations presented in the main text, we have verified that employing swap Monte Carlo moves within a region encompassing $25\%$ of the system allows for the observation of a distinct plateau far from the interface (as exemplified in Fig.~\ref{fig:tx-T01}), indicative of recovered bulk dynamics. We show in Fig.~\ref{fig:rhoswapT008} (left) that the dynamic profiles do not depend on the choice of system size in such conditions.
	
	Regarding the choice of parameters for the swap dynamics, our analysis confirms the independence of fit results upon normalization of the dynamic profile by its bulk value, $\tau/\tau_{\rm smc}$. As depicted in Fig.~\ref{fig:rhoswapT008} (right), even with markedly different relaxation times between two profiles at $T=0.08$, proper normalization reveals an equivalent description of the spreading of mobility regions within the MD domain, thereby confirming the robustness of our methodology (of course in this normalisation, the large $x$ limits cannot be equal).
	
	\textbf{Dynamic profiles in $3d$.}
	
\begin{figure}[h]
		\centering
		\includegraphics{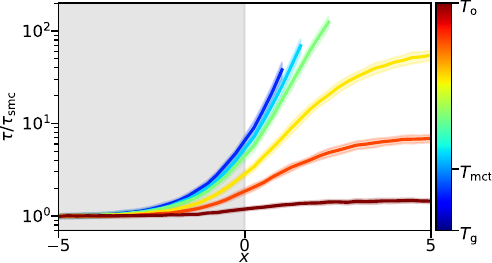}
		\caption{Dynamic profiles showing the evolution of the structural relaxation with $x$, normalised by $\tau_{\rm smc}$ for the $3d$ system. As shown for $2d$ in the main text, dynamic facilitation becomes less efficient at lower $T$}.
		\label{fig:tx-3d}
	\end{figure}

	In addition to Fig.~1 in the main text, we present in Fig.~\ref{fig:tx-3d} the dynamic profiles of the $3d$ simulations, conducted accross temperatures $T \in [0.07,0.2]$. Here, the glass transition temperature is $T_{\rm g} = 0.056$, the mode coupling temperature is $T_{\rm mct} = 0.095$ and the onset temperature is $T_{\rm o} = 0.2$. Our observation reveals again (as in $2d$) a strong decrease in dynamic facilitation efficiency as temperature decreases, and an increasingly subdiffusive motion. The error bars in the $3d$ system are smaller than those in its $2d$ counterpart, owing to the utilization of a greater number of particles.
	
	\hspace{0.5cm}
	
	\textbf{Debye-Waller factor.}
	
	The Debye-Waller factor $\langle u^2 \rangle$ is defined as the amplitude of the short-time vibrational motion~\cite{larini2008,berthier2020}. It is computed from the mean-squared displacement
	\begin{equation}
		\Delta (t) = \frac{1}{N} \sum_i \langle \abs{r_i(t) - r_i(0)} \rangle^2,
	\end{equation}
	where the cage-relative correction of the coordinates has been applied for the $2d$ system~\cite{flenner2015,illing2017}. Specifically, $\langle u^2 \rangle = \Delta (t^*)$, where $t^*$ corresponds to the inflection point of the mean-squared displacement, \textit{i.e.} the minimum of the curve $\dd \log(\langle x^2 \rangle) / \dd t$.
	
	\begin{figure}[h]
		\centering
		\includegraphics{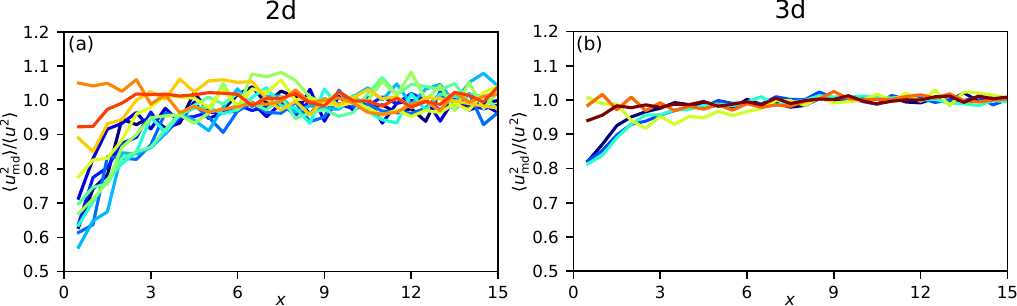}
		\caption{Results of the Debye-Waller factor, normalized by the bulk MD value for different temperatures. The characteristic lengthscale over which one recovers the bulk behavior does not depend on $T$. It is microscopic and largely disconnected from the dynamics.}
		\label{fig:figS2}
	\end{figure}
	
	In Fig.~\ref{fig:figS2} we show the measured profiles of the Debye Waller factor for both $2d$ and $3d$ models. In both cases, we find that the Debye Waller factor is larger near $x=0$, which confirms that the SMC dynamics makes the left side of the simulation dynamically softer even though it is structurally indistinguishable from the right side. We also note that the Debye Waller factor returns to its bulk value at $x>0$ over a microscopic lengthscale which does not grow as temperature is varied over a broad range. This agrees with earlier results in thin polymer films~\cite{sussman2017}. There is therefore a large disconnect between short-time dynamics and elasticity and the long-time dynamic profiles. 
	
	\hspace{0.5cm}
	
	\textbf{Variance of the mobile interface.}
	\begin{figure}[h]
		\centering
		\includegraphics{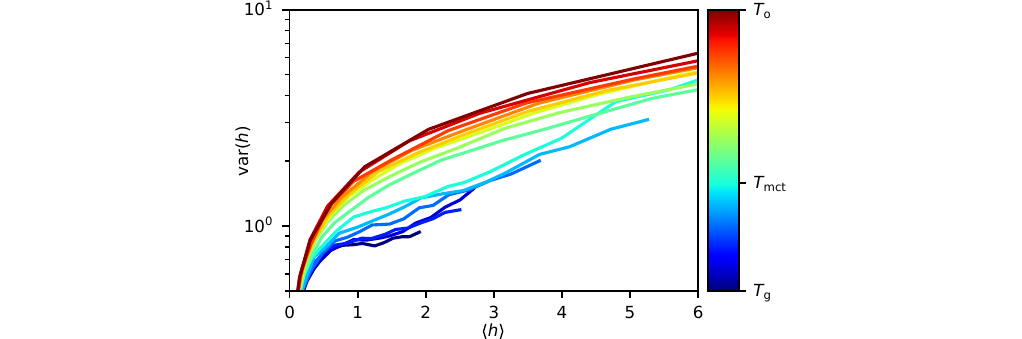}
		\caption{Variance of the mobile interface for $2d$ simulations as a function of the average interface position. The interface roughness decreases when getting closer to $T_{\rm g}$ in the displacement range where the dynamic profiles are described by power law and logarithmic fits, respectively.}
		\label{fig:figS3}
	\end{figure}
	
	One can define an interface between mobile and immobile particles close to the frontier at $x=0$. Then, as dynamic facilitation acts over time, the mobile interface moves to a new position $h(x,t)$. To characterise $h(x,t)$ we follow the procedure described in Ref.~\cite{herrero2023front}, but we distinguish mobile and immobile particles for a threshold $C_B^i=0.7$. The average interface position is defined from the first moment:
	\begin{equation}
		\langle h \rangle (t) = \frac{1}{L}\int_0^L \dd x' h(x',t),
	\end{equation}
	with $L$ the linear box size in the $y$ direction.
	Then, the roughness of the relaxed domains is given by the standard deviation,
	\begin{equation}
		{\rm var}(h) = \sqrt{\langle h^2 \rangle - \langle h \rangle^2},
	\end{equation}
	where
	\begin{equation}
		\langle h^2 \rangle (t) = \frac{1}{L}\int_0^L \dd x' h^2(x',t).
	\end{equation}
	
	In Fig.~\ref{fig:figS3} we represent a parametric plot of the variance ${\rm var}(h)$ against the mean $\langle h \rangle$ for various temperatures. We observe that the rugosity of the interface becomes less pronounced at temperature goes down. 
	
	\hspace{0.5cm}
	
	\textbf{The inhomogeneous swap-MD scheme produces homogeneous structure.}
	
\begin{figure}[h]
		\centering
		\includegraphics{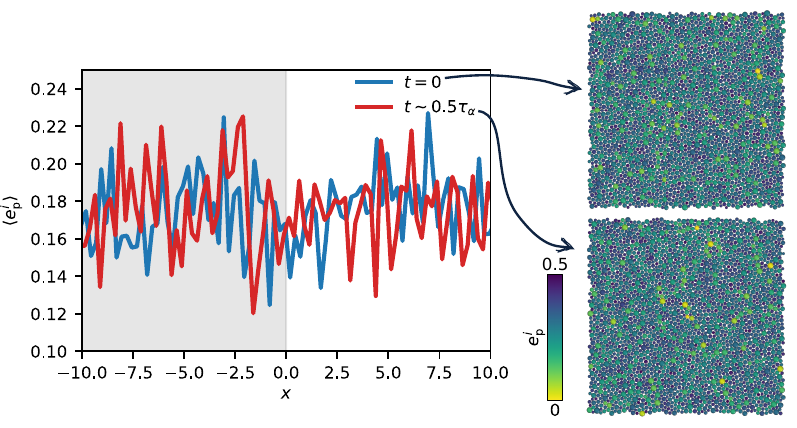}
		\caption{Potential energy per particle at $T=0.095$ for the $2d$ system computed in the inherent state, averaged over bins in the $x$ direction with a size of ${\rm d}x=0.25$. We note that it exhibits equivalent profiles at $t=0$ (initial configuration) and at long times after the hybrid swap-MD simulations have been performed, within both the swap ($x<0$) and MD regions. Consistently, the two system snapshots on the right show a homogeneous structure within the system, where the regions of fast and the slow dynamics are indistinguishable.}
		\label{fig:epx}
	\end{figure}

	To ensure the absence of interfacial effects in the structure caused by our space-dependent swap-MD algorithm, we conducted a thorough verification process to confirm that both the SMC and MD regions exhibit identical structures with no interfacial cost. In Fig.~\ref{fig:epx}, we present the potential energy profile across the SMC-MD interface, as we have done for the dynamic profile representations (e.g. Fig.~\ref{fig:figS1}). This figure depicts the potential energy at $t=0$, before the initiation of hybrid swap-MD simulations, with that observed at long times, where mobility spreads from the SMC to the MD region. Both profiles are visually indistinguishable, as determined through both visual inspection and analysis of potential energy profiles, which remain equivalent within the noise both at the level of the average and the fluctuations.
	
\begin{figure}[h!]
		\centering
		\includegraphics[width=0.6\textwidth]{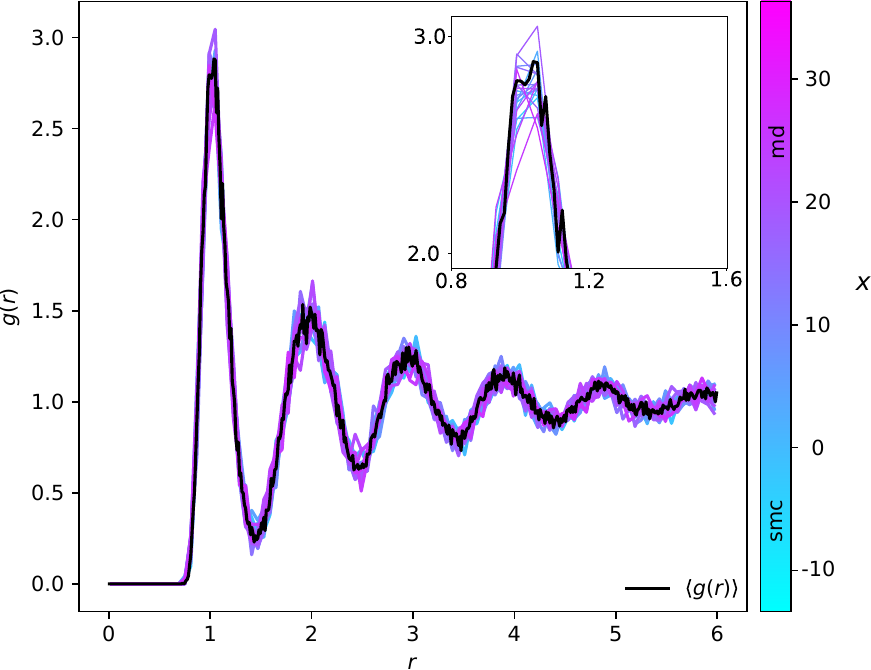}
		\caption{Radial distribution function at $T=0.1$ for the $2d$ system, averaged over bins in the $x$ direction with a size of $\Delta x=2.5$. The system exhibits equivalent profiles at different distances from the dynamic frontier, located at $x=0$. Additionally, all the bin results are consistent with the radial distribution function computed for the full box, shown in black. The inset corresponds to a zoom of the first peak of the $g(r)$ curves, where only statistical noise can be observed with no systematic trend. We obtained equivalent results for all temperatures.}
		\label{fig:gr}
	\end{figure}

	Another test we performed to ensure the uniformity of the structure within the system is the computation of the radial distribution function, $g(r)$, at different distances from the interface, as shown in Fig.~\ref{fig:gr}. To achieve this, we define the partial $g(r)$ at a fixed distance $x$ from the interface at $x=0$, using the normalized distribution of particles $j$ at a distance $r_ij$ from particle $i$, with $i$ located in the bin $[x - \Delta x,x+\Delta x]$. By resolving spatially the pair correlation function, we confirm that all measurements are independent of $x$ within the statistical accuracy. Furthermore, all results are consistent with the ensemble average of $g(r)$ (depicted in black), obtained by averaging over all $i$ particles within the entire box. This confirms that at the structural level, the swap-MD algorithm maintains perfect homogeneity and equilibrium properties of the system.
	
	\hspace{0.5cm}

	\textbf{Relaxation of isolated domains in bulk.}
	
	We wish to confirm that the measured value of the dynamic exponent $z$ obtained with the facilitated front geometry of the present work is consistent with the spatial spreading of mobile regions in bulk.
	
\begin{figure}[h!]
		\centering
		\includegraphics{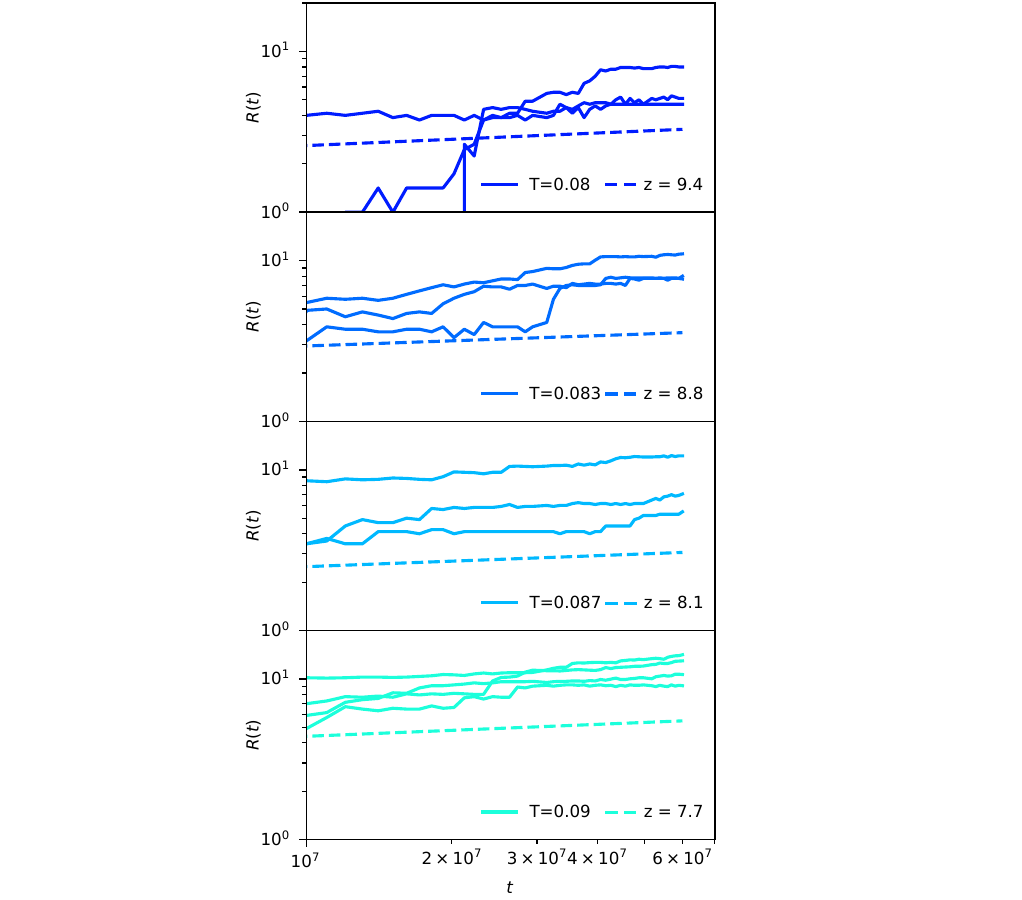}
		\caption{Spatial spreading of selected isolated mobile regions in an independent series of bulk MD simulations for different temperatures. The reported curves show good agreement with a power-law growth with the $z$ exponents obtained from the facilitated front geometry described in the main text (dashed lines).}
		\label{fig:figS4}
	\end{figure}

	To this end, we perform a large number of bulk simulations using MD dynamics at various temperatures. From these runs, we select a subset of isolated mobile regions in bulk MD simulations, for $N=10000$, for four different temperatures, $T=\{0.08,0.083,0.084,0.09\}$ for the $2d$ system. For each selected domain, we measure the time evolution of the number of mobile particles in each domain, $n(t)$, which we convert into a lengthscale $R(t) = \sqrt{n(t)}$. The time dependence of $R(t)$ should then obey:   
	\begin{equation}
		R(t) \propto t^{1/z(T)}.
	\end{equation}
	Figure~\ref{fig:figS4} shows that, within the large domain-to-domain fluctuations, a good agreement is obtained between the averaged $z$ values deduced from the macroscopic interface geometry used in the main text, and the time evolution of individual domains.
	
	\hspace{0.5cm}

	\textbf{Comparison with Keys et al.}
	
	In Ref.~\citenum{keys2011} Keys and collaborators develop a comprehensive analysis of dynamic facilitation. We detail here how to obtain the dynamic exponent $z(T)$ and the lengthscale $\xi_4$, corresponding to Eq.~6 of the main text, from their data.
	
	The relaxation picture in Ref.~\citenum{keys2011} consists in the propagation in space of elementary excitations. Then, the equilibrium concentration of excitations follows a Boltzmann temperature dependence in the form:
	\begin{equation}
		c(a,T) = \exp(- \frac{J(a)}{\Tilde{T}} ) \qquad {\rm with} \quad \Tilde{T} = \qty(\frac{1}{T} - \frac{1}{T_{\rm o}})^{-1},
		\label{eq:concentration}
	\end{equation}
	where $T_{\rm o}$ is the onset temperature and $a$ a characteristic length. Then, the characteristic distance between excitations of size $a$ is given by $\ell_a/a \sim [c(a,T)]^{-1/d_{\rm f}}$, with $d_{\rm f}$ the fractal dimension. Combining both expressions, and supposing that the size of the elementary excitation is on the order of the molecular diameter $\sigma$ (as assumed by the authors), we obtain:
	\begin{equation}
		\xi_{4} = \ell_\sigma \sim \exp( \frac{J_\sigma}{d_{\rm f} \Tilde{T}}).
		\label{eq:xi}
	\end{equation}
	
	To obtain the dynamic exponent $z$, we start from their consideration of the logarithmic growth of the energy barrier $J_a$ with $a$, $J_a - J_{a'} = \gamma J_\sigma \ln(a/a')$, where $\gamma$ is a dimensionless factor on the order of the unity and $J_\sigma$ a material-dependent reference energy. One can rewrite this expression in terms of concentrations following Eq.~\ref{eq:concentration}:
	\begin{equation}
		c(a',T) = c(a) \qty(\frac{a'}{a})^{-\gamma J_\sigma/\Tilde{T}}.
	\end{equation}
	It is then clear from this expression that the ``growth'' of a domain of size $a$ to a larger domain of size $a'$, follows a power-law behavior with exponent $\gamma J_\sigma /\Tilde{T}$. If we compare it with the power law fit in the main text, this growth is equivalent to the dynamic spreading of the domains with $z(T)$, and thus:
	\begin{equation}
		z(T) = \frac{ \gamma J_\sigma }{ \Tilde{T} }.
		\label{eq:z}
	\end{equation}
	In other words, the logarithmic growth of the energy barrier is equivalent to the subdiffusive growth of relaxed domains. Note that following these two expressions, Eq.~\ref{eq:xi} for $\xi_4$ and Eq.~\ref{eq:z} for $z$, and substituting them in Eq.~(6) of the main text, we obtain:
	\begin{equation}
		\tau_\alpha(\Tilde{T}) \sim \tau_{\rm o} \exp( \frac{\gamma J_\sigma^2}{d_{\rm f} } \frac{1}{\Tilde{T}^2} ),
	\end{equation}
	and thus recovering the full $\tau_\alpha$ relaxation behavior described in Ref.~\citenum{keys2011}.

%\bibliography{cecilia.bib}
\input{si.bbl}

%% file: main.bbl

%% file: si.bbl

%% file: main.bbl
\begin{thebibliography}{57}%
\makeatletter
\providecommand \@ifxundefined [1]{%
 \@ifx{#1\undefined}
}%
\providecommand \@ifnum [1]{%
 \ifnum #1\expandafter \@firstoftwo
 \else \expandafter \@secondoftwo
 \fi
}%
\providecommand \@ifx [1]{%
 \ifx #1\expandafter \@firstoftwo
 \else \expandafter \@secondoftwo
 \fi
}%
\providecommand \natexlab [1]{#1}%
\providecommand \enquote  [1]{``#1''}%
\providecommand \bibnamefont  [1]{#1}%
\providecommand \bibfnamefont [1]{#1}%
\providecommand \citenamefont [1]{#1}%
\providecommand \href@noop [0]{\@secondoftwo}%
\providecommand \href [0]{\begingroup \@sanitize@url \@href}%
\providecommand \@href[1]{\@@startlink{#1}\@@href}%
\providecommand \@@href[1]{\endgroup#1\@@endlink}%
\providecommand \@sanitize@url [0]{\catcode `\\12\catcode `\$12\catcode
  `\&12\catcode `\#12\catcode `\^12\catcode `\_12\catcode `\%12\relax}%
\providecommand \@@startlink[1]{}%
\providecommand \@@endlink[0]{}%
\providecommand \url  [0]{\begingroup\@sanitize@url \@url }%
\providecommand \@url [1]{\endgroup\@href {#1}{\urlprefix }}%
\providecommand \urlprefix  [0]{URL }%
\providecommand \Eprint [0]{\href }%
\providecommand \doibase [0]{https://doi.org/}%
\providecommand \selectlanguage [0]{\@gobble}%
\providecommand \bibinfo  [0]{\@secondoftwo}%
\providecommand \bibfield  [0]{\@secondoftwo}%
\providecommand \translation [1]{[#1]}%
\providecommand \BibitemOpen [0]{}%
\providecommand \bibitemStop [0]{}%
\providecommand \bibitemNoStop [0]{.\EOS\space}%
\providecommand \EOS [0]{\spacefactor3000\relax}%
\providecommand \BibitemShut  [1]{\csname bibitem#1\endcsname}%
\let\auto@bib@innerbib\@empty
%</preamble>
\bibitem [{\citenamefont {Ediger}\ \emph {et~al.}(1996)\citenamefont {Ediger},
  \citenamefont {Angell},\ and\ \citenamefont {Nagel}}]{ediger1996}%
  \BibitemOpen
  \bibfield  {author} {\bibinfo {author} {\bibfnamefont {M.~D.}\ \bibnamefont
  {Ediger}}, \bibinfo {author} {\bibfnamefont {C.~A.}\ \bibnamefont {Angell}},\
  and\ \bibinfo {author} {\bibfnamefont {S.~R.}\ \bibnamefont {Nagel}},\
  }\bibfield  {title} {\bibinfo {title} {Supercooled liquids and glasses},\
  }\href {https://doi.org/10.1021/jp953538d} {\bibfield  {journal} {\bibinfo
  {journal} {The Journal of Physical Chemistry}\ }\textbf {\bibinfo {volume}
  {100}},\ \bibinfo {pages} {13200} (\bibinfo {year} {1996})}\BibitemShut
  {NoStop}%
\bibitem [{\citenamefont {Ediger}(2000)}]{ediger2000spatially}%
  \BibitemOpen
  \bibfield  {author} {\bibinfo {author} {\bibfnamefont {M.~D.}\ \bibnamefont
  {Ediger}},\ }\bibfield  {title} {\bibinfo {title} {Spatially heterogeneous
  dynamics in supercooled liquids},\ }\href
  {https://doi.org/10.1146/annurev.physchem.51.1.99} {\bibfield  {journal}
  {\bibinfo  {journal} {Annual Review of Physical Chemistry}\ }\textbf
  {\bibinfo {volume} {51}},\ \bibinfo {pages} {99} (\bibinfo {year}
  {2000})}\BibitemShut {NoStop}%
\bibitem [{\citenamefont {Debenedetti}\ and\ \citenamefont
  {Stillinger}(2001)}]{debenedetti2001supercooled}%
  \BibitemOpen
  \bibfield  {author} {\bibinfo {author} {\bibfnamefont {P.~G.}\ \bibnamefont
  {Debenedetti}}\ and\ \bibinfo {author} {\bibfnamefont {F.~H.}\ \bibnamefont
  {Stillinger}},\ }\bibfield  {title} {\bibinfo {title} {Supercooled liquids
  and the glass transition},\ }\href {https://doi.org/10.1038/35065704}
  {\bibfield  {journal} {\bibinfo  {journal} {Nature}\ }\textbf {\bibinfo
  {volume} {410}},\ \bibinfo {pages} {259} (\bibinfo {year}
  {2001})}\BibitemShut {NoStop}%
\bibitem [{\citenamefont {Berthier}\ and\ \citenamefont
  {Biroli}(2011)}]{berthier2011}%
  \BibitemOpen
  \bibfield  {author} {\bibinfo {author} {\bibfnamefont {L.}~\bibnamefont
  {Berthier}}\ and\ \bibinfo {author} {\bibfnamefont {G.}~\bibnamefont
  {Biroli}},\ }\bibfield  {title} {\bibinfo {title} {Theoretical perspective on
  the glass transition and amorphous materials},\ }\href
  {https://doi.org/10.1103/RevModPhys.83.587} {\bibfield  {journal} {\bibinfo
  {journal} {Rev. Mod. Phys.}\ }\textbf {\bibinfo {volume} {83}},\ \bibinfo
  {pages} {587} (\bibinfo {year} {2011})}\BibitemShut {NoStop}%
\bibitem [{\citenamefont {Palmer}\ \emph {et~al.}(1984)\citenamefont {Palmer},
  \citenamefont {Stein}, \citenamefont {Abrahams},\ and\ \citenamefont
  {Anderson}}]{palmer1984}%
  \BibitemOpen
  \bibfield  {author} {\bibinfo {author} {\bibfnamefont {R.~G.}\ \bibnamefont
  {Palmer}}, \bibinfo {author} {\bibfnamefont {D.~L.}\ \bibnamefont {Stein}},
  \bibinfo {author} {\bibfnamefont {E.}~\bibnamefont {Abrahams}},\ and\
  \bibinfo {author} {\bibfnamefont {P.~W.}\ \bibnamefont {Anderson}},\
  }\bibfield  {title} {\bibinfo {title} {Models of hierarchically constrained
  dynamics for glassy relaxation},\ }\href
  {https://doi.org/10.1103/PhysRevLett.53.958} {\bibfield  {journal} {\bibinfo
  {journal} {Phys. Rev. Lett.}\ }\textbf {\bibinfo {volume} {53}},\ \bibinfo
  {pages} {958} (\bibinfo {year} {1984})}\BibitemShut {NoStop}%
\bibitem [{\citenamefont {Fredrickson}\ and\ \citenamefont
  {Andersen}(1984)}]{fredrickson1984kinetic}%
  \BibitemOpen
  \bibfield  {author} {\bibinfo {author} {\bibfnamefont {G.~H.}\ \bibnamefont
  {Fredrickson}}\ and\ \bibinfo {author} {\bibfnamefont {H.~C.}\ \bibnamefont
  {Andersen}},\ }\bibfield  {title} {\bibinfo {title} {Kinetic ising model of
  the glass transition},\ }\href {https://doi.org/10.1103/PhysRevLett.53.1244}
  {\bibfield  {journal} {\bibinfo  {journal} {Phys. Rev. Lett.}\ }\textbf
  {\bibinfo {volume} {53}},\ \bibinfo {pages} {1244} (\bibinfo {year}
  {1984})}\BibitemShut {NoStop}%
\bibitem [{\citenamefont {J{\"a}ckle}\ and\ \citenamefont
  {Eisinger}(1991)}]{jackle1991}%
  \BibitemOpen
  \bibfield  {author} {\bibinfo {author} {\bibfnamefont {J.}~\bibnamefont
  {J{\"a}ckle}}\ and\ \bibinfo {author} {\bibfnamefont {S.}~\bibnamefont
  {Eisinger}},\ }\bibfield  {title} {\bibinfo {title} {A hierarchically
  constrained kinetic ising model},\ }\href
  {https://link.springer.com/article/10.1007/BF01453764#citeas} {\bibfield
  {journal} {\bibinfo  {journal} {Zeitschrift f{\"u}r physik B condensed
  matter}\ }\textbf {\bibinfo {volume} {84}},\ \bibinfo {pages} {115} (\bibinfo
  {year} {1991})}\BibitemShut {NoStop}%
\bibitem [{\citenamefont {Chandler}\ and\ \citenamefont
  {Garrahan}(2010)}]{chandler2010dynamics}%
  \BibitemOpen
  \bibfield  {author} {\bibinfo {author} {\bibfnamefont {D.}~\bibnamefont
  {Chandler}}\ and\ \bibinfo {author} {\bibfnamefont {J.~P.}\ \bibnamefont
  {Garrahan}},\ }\bibfield  {title} {\bibinfo {title} {Dynamics on the way to
  forming glass: Bubbles in space-time},\ }\href
  {https://doi.org/10.1146/annurev.physchem.040808.090405} {\bibfield
  {journal} {\bibinfo  {journal} {Annual Review of Physical Chemistry}\
  }\textbf {\bibinfo {volume} {61}},\ \bibinfo {pages} {191} (\bibinfo {year}
  {2010})}\BibitemShut {NoStop}%
\bibitem [{\citenamefont {Xia}\ and\ \citenamefont
  {Wolynes}(2001)}]{xia2001microscopic}%
  \BibitemOpen
  \bibfield  {author} {\bibinfo {author} {\bibfnamefont {X.}~\bibnamefont
  {Xia}}\ and\ \bibinfo {author} {\bibfnamefont {P.~G.}\ \bibnamefont
  {Wolynes}},\ }\bibfield  {title} {\bibinfo {title} {Microscopic theory of
  heterogeneity and nonexponential relaxations in supercooled liquids},\ }\href
  {https://doi.org/10.1103/PhysRevLett.86.5526} {\bibfield  {journal} {\bibinfo
   {journal} {Phys. Rev. Lett.}\ }\textbf {\bibinfo {volume} {86}},\ \bibinfo
  {pages} {5526} (\bibinfo {year} {2001})}\BibitemShut {NoStop}%
\bibitem [{\citenamefont {Mirigian}\ and\ \citenamefont
  {Schweizer}(2013)}]{mirigian2013unified}%
  \BibitemOpen
  \bibfield  {author} {\bibinfo {author} {\bibfnamefont {S.}~\bibnamefont
  {Mirigian}}\ and\ \bibinfo {author} {\bibfnamefont {K.~S.}\ \bibnamefont
  {Schweizer}},\ }\bibfield  {title} {\bibinfo {title} {Unified theory of
  activated relaxation in liquids over 14 decades in time},\ }\href
  {https://doi.org/10.1021/jz4018943} {\bibfield  {journal} {\bibinfo
  {journal} {The Journal of Physical Chemistry Letters}\ }\textbf {\bibinfo
  {volume} {4}},\ \bibinfo {pages} {3648} (\bibinfo {year} {2013})}\BibitemShut
  {NoStop}%
\bibitem [{\citenamefont {Schoenholz}\ \emph {et~al.}(2016)\citenamefont
  {Schoenholz}, \citenamefont {Cubuk}, \citenamefont {Sussman},\ and\
  \citenamefont {Liu}}]{schoenholz2016structural}%
  \BibitemOpen
  \bibfield  {author} {\bibinfo {author} {\bibfnamefont {S.~S.}\ \bibnamefont
  {Schoenholz}}, \bibinfo {author} {\bibfnamefont {E.~D.}\ \bibnamefont
  {Cubuk}}, \bibinfo {author} {\bibfnamefont {D.~M.}\ \bibnamefont {Sussman}},\
  and\ \bibinfo {author} {\bibfnamefont {A.~J.}\ \bibnamefont {Liu}},\
  }\bibfield  {title} {\bibinfo {title} {A structural approach to relaxation in
  glassy liquids},\ }\href {https://doi.org/https://doi.org/10.1038/nphys3644}
  {\bibfield  {journal} {\bibinfo  {journal} {Nature Physics}\ }\textbf
  {\bibinfo {volume} {12}},\ \bibinfo {pages} {469} (\bibinfo {year}
  {2016})}\BibitemShut {NoStop}%
\bibitem [{\citenamefont {Speck}(2019)}]{speck2019}%
  \BibitemOpen
  \bibfield  {author} {\bibinfo {author} {\bibfnamefont {T.}~\bibnamefont
  {Speck}},\ }\bibfield  {title} {\bibinfo {title} {Dynamic facilitation
  theory: a statistical mechanics approach to dynamic arrest},\ }\href
  {https://iopscience.iop.org/article/10.1088/1742-5468/ab2ace/meta?casa_token=2LwehWtaXIUAAAAA:8dYqAmlFul-4hh9Rt1v9Vsg17rjrMfj8pAUe90ilKpgQXqCz5UiXJMUR3P1tkCKxmR4G_-Yr0ejg}
  {\bibfield  {journal} {\bibinfo  {journal} {Journal of Statistical Mechanics:
  Theory and Experiment}\ }\textbf {\bibinfo {volume} {2019}},\ \bibinfo
  {pages} {084015} (\bibinfo {year} {2019})}\BibitemShut {NoStop}%
\bibitem [{\citenamefont {Ozawa}\ and\ \citenamefont
  {Biroli}(2023)}]{ozawa2023}%
  \BibitemOpen
  \bibfield  {author} {\bibinfo {author} {\bibfnamefont {M.}~\bibnamefont
  {Ozawa}}\ and\ \bibinfo {author} {\bibfnamefont {G.}~\bibnamefont {Biroli}},\
  }\bibfield  {title} {\bibinfo {title} {Elasticity, facilitation, and dynamic
  heterogeneity in glass-forming liquids},\ }\href
  {https://doi.org/10.1103/PhysRevLett.130.138201} {\bibfield  {journal}
  {\bibinfo  {journal} {Physical Review Letters}\ }\textbf {\bibinfo {volume}
  {130}},\ \bibinfo {pages} {138201} (\bibinfo {year} {2023})}\BibitemShut
  {NoStop}%
\bibitem [{\citenamefont {Tahaei}\ \emph {et~al.}(2023)\citenamefont {Tahaei},
  \citenamefont {Biroli}, \citenamefont {Ozawa}, \citenamefont
  {Popovi\ifmmode~\acute{c}\else \'{c}\fi{}},\ and\ \citenamefont
  {Wyart}}]{tahaei2023}%
  \BibitemOpen
  \bibfield  {author} {\bibinfo {author} {\bibfnamefont {A.}~\bibnamefont
  {Tahaei}}, \bibinfo {author} {\bibfnamefont {G.}~\bibnamefont {Biroli}},
  \bibinfo {author} {\bibfnamefont {M.}~\bibnamefont {Ozawa}}, \bibinfo
  {author} {\bibfnamefont {M.}~\bibnamefont {Popovi\ifmmode~\acute{c}\else
  \'{c}\fi{}}},\ and\ \bibinfo {author} {\bibfnamefont {M.}~\bibnamefont
  {Wyart}},\ }\bibfield  {title} {\bibinfo {title} {Scaling description of
  dynamical heterogeneity and avalanches of relaxation in glass-forming
  liquids},\ }\href {https://doi.org/10.1103/PhysRevX.13.031034} {\bibfield
  {journal} {\bibinfo  {journal} {Phys. Rev. X}\ }\textbf {\bibinfo {volume}
  {13}},\ \bibinfo {pages} {031034} (\bibinfo {year} {2023})}\BibitemShut
  {NoStop}%
\bibitem [{\citenamefont {Ghanekarade}\ \emph {et~al.}(2023)\citenamefont
  {Ghanekarade}, \citenamefont {Phan}, \citenamefont {Schweizer},\ and\
  \citenamefont {Simmons}}]{ghanekarade2023}%
  \BibitemOpen
  \bibfield  {author} {\bibinfo {author} {\bibfnamefont {A.}~\bibnamefont
  {Ghanekarade}}, \bibinfo {author} {\bibfnamefont {A.~D.}\ \bibnamefont
  {Phan}}, \bibinfo {author} {\bibfnamefont {K.~S.}\ \bibnamefont
  {Schweizer}},\ and\ \bibinfo {author} {\bibfnamefont {D.~S.}\ \bibnamefont
  {Simmons}},\ }\bibfield  {title} {\bibinfo {title} {Signature of collective
  elastic glass physics in surface-induced long-range tails in dynamical
  gradients},\ }\href {https://www.nature.com/articles/s41567-023-01995-8}
  {\bibfield  {journal} {\bibinfo  {journal} {Nature Physics}\ }\textbf
  {\bibinfo {volume} {19}},\ \bibinfo {pages} {800} (\bibinfo {year}
  {2023})}\BibitemShut {NoStop}%
\bibitem [{\citenamefont {Vogel}\ and\ \citenamefont
  {Glotzer}(2004)}]{vogel2004spatially}%
  \BibitemOpen
  \bibfield  {author} {\bibinfo {author} {\bibfnamefont {M.}~\bibnamefont
  {Vogel}}\ and\ \bibinfo {author} {\bibfnamefont {S.~C.}\ \bibnamefont
  {Glotzer}},\ }\bibfield  {title} {\bibinfo {title} {Spatially heterogeneous
  dynamics and dynamic facilitation in a model of viscous silica},\ }\href
  {https://doi.org/10.1103/PhysRevLett.92.255901} {\bibfield  {journal}
  {\bibinfo  {journal} {Phys. Rev. Lett.}\ }\textbf {\bibinfo {volume} {92}},\
  \bibinfo {pages} {255901} (\bibinfo {year} {2004})}\BibitemShut {NoStop}%
\bibitem [{\citenamefont {Bergroth}\ \emph {et~al.}(2005)\citenamefont
  {Bergroth}, \citenamefont {Vogel},\ and\ \citenamefont
  {Glotzer}}]{bergroth2005examination}%
  \BibitemOpen
  \bibfield  {author} {\bibinfo {author} {\bibfnamefont {M.~N.~J.}\
  \bibnamefont {Bergroth}}, \bibinfo {author} {\bibfnamefont {M.}~\bibnamefont
  {Vogel}},\ and\ \bibinfo {author} {\bibfnamefont {S.~C.}\ \bibnamefont
  {Glotzer}},\ }\bibfield  {title} {\bibinfo {title} {Examination of dynamic
  facilitation in molecular dynamics simulations of glass-forming liquids},\
  }\href {https://doi.org/10.1021/jp0447946} {\bibfield  {journal} {\bibinfo
  {journal} {The Journal of Physical Chemistry B}\ }\textbf {\bibinfo {volume}
  {109}},\ \bibinfo {pages} {6748} (\bibinfo {year} {2005})}\BibitemShut
  {NoStop}%
\bibitem [{\citenamefont {Keys}\ \emph {et~al.}(2011)\citenamefont {Keys},
  \citenamefont {Hedges}, \citenamefont {Garrahan}, \citenamefont {Glotzer},\
  and\ \citenamefont {Chandler}}]{keys2011}%
  \BibitemOpen
  \bibfield  {author} {\bibinfo {author} {\bibfnamefont {A.~S.}\ \bibnamefont
  {Keys}}, \bibinfo {author} {\bibfnamefont {L.~O.}\ \bibnamefont {Hedges}},
  \bibinfo {author} {\bibfnamefont {J.~P.}\ \bibnamefont {Garrahan}}, \bibinfo
  {author} {\bibfnamefont {S.~C.}\ \bibnamefont {Glotzer}},\ and\ \bibinfo
  {author} {\bibfnamefont {D.}~\bibnamefont {Chandler}},\ }\bibfield  {title}
  {\bibinfo {title} {Excitations are localized and relaxation is hierarchical
  in glass-forming liquids},\ }\href
  {https://doi.org/10.1103/PhysRevX.1.021013} {\bibfield  {journal} {\bibinfo
  {journal} {Physical Review X}\ }\textbf {\bibinfo {volume} {1}},\ \bibinfo
  {pages} {021013} (\bibinfo {year} {2011})}\BibitemShut {NoStop}%
\bibitem [{\citenamefont {Gokhale}\ \emph {et~al.}(2014)\citenamefont
  {Gokhale}, \citenamefont {Hima~Nagamanasa}, \citenamefont {Ganapathy},\ and\
  \citenamefont {Sood}}]{gokhale2014growing}%
  \BibitemOpen
  \bibfield  {author} {\bibinfo {author} {\bibfnamefont {S.}~\bibnamefont
  {Gokhale}}, \bibinfo {author} {\bibfnamefont {K.}~\bibnamefont
  {Hima~Nagamanasa}}, \bibinfo {author} {\bibfnamefont {R.}~\bibnamefont
  {Ganapathy}},\ and\ \bibinfo {author} {\bibfnamefont {A.}~\bibnamefont
  {Sood}},\ }\bibfield  {title} {\bibinfo {title} {Growing dynamical
  facilitation on approaching the random pinning colloidal glass transition},\
  }\href {https://doi.org/https://doi.org/10.1038/ncomms5685} {\bibfield
  {journal} {\bibinfo  {journal} {Nature communications}\ }\textbf {\bibinfo
  {volume} {5}},\ \bibinfo {pages} {4685} (\bibinfo {year} {2014})}\BibitemShut
  {NoStop}%
\bibitem [{\citenamefont {Chacko}\ \emph {et~al.}(2021)\citenamefont {Chacko},
  \citenamefont {Landes}, \citenamefont {Biroli}, \citenamefont {Dauchot},
  \citenamefont {Liu},\ and\ \citenamefont {Reichman}}]{chacko2021}%
  \BibitemOpen
  \bibfield  {author} {\bibinfo {author} {\bibfnamefont {R.~N.}\ \bibnamefont
  {Chacko}}, \bibinfo {author} {\bibfnamefont {F.~P.}\ \bibnamefont {Landes}},
  \bibinfo {author} {\bibfnamefont {G.}~\bibnamefont {Biroli}}, \bibinfo
  {author} {\bibfnamefont {O.}~\bibnamefont {Dauchot}}, \bibinfo {author}
  {\bibfnamefont {A.~J.}\ \bibnamefont {Liu}},\ and\ \bibinfo {author}
  {\bibfnamefont {D.~R.}\ \bibnamefont {Reichman}},\ }\bibfield  {title}
  {\bibinfo {title} {Elastoplasticity mediates dynamical heterogeneity below
  the mode coupling temperature},\ }\href
  {https://doi.org/10.1103/PhysRevLett.127.048002} {\bibfield  {journal}
  {\bibinfo  {journal} {Physical Review Letters}\ }\textbf {\bibinfo {volume}
  {127}},\ \bibinfo {pages} {048002} (\bibinfo {year} {2021})}\BibitemShut
  {NoStop}%
\bibitem [{\citenamefont {Isobe}\ \emph {et~al.}(2016)\citenamefont {Isobe},
  \citenamefont {Keys}, \citenamefont {Chandler},\ and\ \citenamefont
  {Garrahan}}]{isobe2016}%
  \BibitemOpen
  \bibfield  {author} {\bibinfo {author} {\bibfnamefont {M.}~\bibnamefont
  {Isobe}}, \bibinfo {author} {\bibfnamefont {A.~S.}\ \bibnamefont {Keys}},
  \bibinfo {author} {\bibfnamefont {D.}~\bibnamefont {Chandler}},\ and\
  \bibinfo {author} {\bibfnamefont {J.~P.}\ \bibnamefont {Garrahan}},\
  }\bibfield  {title} {\bibinfo {title} {Applicability of dynamic facilitation
  theory to binary hard disk systems},\ }\href
  {https://journals.aps.org/prl/abstract/10.1103/PhysRevLett.117.145701}
  {\bibfield  {journal} {\bibinfo  {journal} {Physical Review Letters}\
  }\textbf {\bibinfo {volume} {117}},\ \bibinfo {pages} {145701} (\bibinfo
  {year} {2016})}\BibitemShut {NoStop}%
\bibitem [{\citenamefont {Hasyim}\ and\ \citenamefont
  {Mandadapu}(2021)}]{hasyim2021theory}%
  \BibitemOpen
  \bibfield  {author} {\bibinfo {author} {\bibfnamefont {M.~R.}\ \bibnamefont
  {Hasyim}}\ and\ \bibinfo {author} {\bibfnamefont {K.~K.}\ \bibnamefont
  {Mandadapu}},\ }\bibfield  {title} {\bibinfo {title} {{A theory of localized
  excitations in supercooled liquids}},\ }\href
  {https://doi.org/10.1063/5.0056303} {\bibfield  {journal} {\bibinfo
  {journal} {The Journal of Chemical Physics}\ }\textbf {\bibinfo {volume}
  {155}},\ \bibinfo {pages} {044504} (\bibinfo {year} {2021})}\BibitemShut
  {NoStop}%
\bibitem [{\citenamefont {Sollich}\ and\ \citenamefont
  {Evans}(1999)}]{sollich1999glassy}%
  \BibitemOpen
  \bibfield  {author} {\bibinfo {author} {\bibfnamefont {P.}~\bibnamefont
  {Sollich}}\ and\ \bibinfo {author} {\bibfnamefont {M.~R.}\ \bibnamefont
  {Evans}},\ }\bibfield  {title} {\bibinfo {title} {Glassy time-scale
  divergence and anomalous coarsening in a kinetically constrained spin
  chain},\ }\href {https://doi.org/10.1103/PhysRevLett.83.3238} {\bibfield
  {journal} {\bibinfo  {journal} {Phys. Rev. Lett.}\ }\textbf {\bibinfo
  {volume} {83}},\ \bibinfo {pages} {3238} (\bibinfo {year}
  {1999})}\BibitemShut {NoStop}%
\bibitem [{\citenamefont {Berthier}\ and\ \citenamefont
  {Garrahan}(2005)}]{berthier2005numerical}%
  \BibitemOpen
  \bibfield  {author} {\bibinfo {author} {\bibfnamefont {L.}~\bibnamefont
  {Berthier}}\ and\ \bibinfo {author} {\bibfnamefont {J.~P.}\ \bibnamefont
  {Garrahan}},\ }\bibfield  {title} {\bibinfo {title} {Numerical study of a
  fragile three-dimensional kinetically constrained model},\ }\href
  {https://doi.org/10.1021/jp045491e} {\bibfield  {journal} {\bibinfo
  {journal} {The Journal of Physical Chemistry B}\ }\textbf {\bibinfo {volume}
  {109}},\ \bibinfo {pages} {3578} (\bibinfo {year} {2005})}\BibitemShut
  {NoStop}%
\bibitem [{\citenamefont {Donati}\ \emph {et~al.}(1999)\citenamefont {Donati},
  \citenamefont {Glotzer},\ and\ \citenamefont {Poole}}]{donati1999}%
  \BibitemOpen
  \bibfield  {author} {\bibinfo {author} {\bibfnamefont {C.}~\bibnamefont
  {Donati}}, \bibinfo {author} {\bibfnamefont {S.~C.}\ \bibnamefont
  {Glotzer}},\ and\ \bibinfo {author} {\bibfnamefont {P.~H.}\ \bibnamefont
  {Poole}},\ }\bibfield  {title} {\bibinfo {title} {Growing spatial
  correlations of particle displacements in a simulated liquid on cooling
  toward the glass transition},\ }\href
  {https://doi.org/10.1103/PhysRevLett.82.5064} {\bibfield  {journal} {\bibinfo
   {journal} {Phys. Rev. Lett.}\ }\textbf {\bibinfo {volume} {82}},\ \bibinfo
  {pages} {5064} (\bibinfo {year} {1999})}\BibitemShut {NoStop}%
\bibitem [{\citenamefont {Lačević}\ \emph {et~al.}(2003)\citenamefont
  {Lačević}, \citenamefont {Starr}, \citenamefont {Schrøder},\ and\
  \citenamefont {Glotzer}}]{lacevic2003}%
  \BibitemOpen
  \bibfield  {author} {\bibinfo {author} {\bibfnamefont {N.}~\bibnamefont
  {Lačević}}, \bibinfo {author} {\bibfnamefont {F.~W.}\ \bibnamefont
  {Starr}}, \bibinfo {author} {\bibfnamefont {T.~B.}\ \bibnamefont
  {Schrøder}},\ and\ \bibinfo {author} {\bibfnamefont {S.~C.}\ \bibnamefont
  {Glotzer}},\ }\bibfield  {title} {\bibinfo {title} {{Spatially heterogeneous
  dynamics investigated via a time-dependent four-point density correlation
  function}},\ }\href {https://doi.org/10.1063/1.1605094} {\bibfield  {journal}
  {\bibinfo  {journal} {The Journal of Chemical Physics}\ }\textbf {\bibinfo
  {volume} {119}},\ \bibinfo {pages} {7372} (\bibinfo {year}
  {2003})}\BibitemShut {NoStop}%
\bibitem [{\citenamefont {Toninelli}\ \emph {et~al.}(2005)\citenamefont
  {Toninelli}, \citenamefont {Wyart}, \citenamefont {Berthier}, \citenamefont
  {Biroli},\ and\ \citenamefont {Bouchaud}}]{toninelli2005}%
  \BibitemOpen
  \bibfield  {author} {\bibinfo {author} {\bibfnamefont {C.}~\bibnamefont
  {Toninelli}}, \bibinfo {author} {\bibfnamefont {M.}~\bibnamefont {Wyart}},
  \bibinfo {author} {\bibfnamefont {L.}~\bibnamefont {Berthier}}, \bibinfo
  {author} {\bibfnamefont {G.}~\bibnamefont {Biroli}},\ and\ \bibinfo {author}
  {\bibfnamefont {J.-P.}\ \bibnamefont {Bouchaud}},\ }\bibfield  {title}
  {\bibinfo {title} {Dynamical susceptibility of glass formers: Contrasting the
  predictions of theoretical scenarios},\ }\href
  {https://journals.aps.org/pre/abstract/10.1103/PhysRevE.71.041505} {\bibfield
   {journal} {\bibinfo  {journal} {Physical Review E}\ }\textbf {\bibinfo
  {volume} {71}},\ \bibinfo {pages} {041505} (\bibinfo {year}
  {2005})}\BibitemShut {NoStop}%
\bibitem [{\citenamefont {Karmakar}\ \emph {et~al.}(2010)\citenamefont
  {Karmakar}, \citenamefont {Dasgupta},\ and\ \citenamefont
  {Sastry}}]{karmakar2010}%
  \BibitemOpen
  \bibfield  {author} {\bibinfo {author} {\bibfnamefont {S.}~\bibnamefont
  {Karmakar}}, \bibinfo {author} {\bibfnamefont {C.}~\bibnamefont {Dasgupta}},\
  and\ \bibinfo {author} {\bibfnamefont {S.}~\bibnamefont {Sastry}},\
  }\bibfield  {title} {\bibinfo {title} {Analysis of dynamic heterogeneity in a
  glass former from the spatial correlations of mobility},\ }\href
  {https://doi.org/10.1103/PhysRevLett.105.015701} {\bibfield  {journal}
  {\bibinfo  {journal} {Phys. Rev. Lett.}\ }\textbf {\bibinfo {volume} {105}},\
  \bibinfo {pages} {015701} (\bibinfo {year} {2010})}\BibitemShut {NoStop}%
\bibitem [{\citenamefont {Berthier}\ \emph {et~al.}(2011)\citenamefont
  {Berthier}, \citenamefont {Biroli}, \citenamefont {Bouchaud}, \citenamefont
  {Cipelletti},\ and\ \citenamefont {van Saarloos}}]{berthier2011dynamical}%
  \BibitemOpen
  \bibfield  {author} {\bibinfo {author} {\bibfnamefont {L.}~\bibnamefont
  {Berthier}}, \bibinfo {author} {\bibfnamefont {G.}~\bibnamefont {Biroli}},
  \bibinfo {author} {\bibfnamefont {J.-P.}\ \bibnamefont {Bouchaud}}, \bibinfo
  {author} {\bibfnamefont {L.}~\bibnamefont {Cipelletti}},\ and\ \bibinfo
  {author} {\bibfnamefont {W.}~\bibnamefont {van Saarloos}},\ }\href
  {https://doi.org/10.1093/acprof:oso/9780199691470.001.0001} {\emph {\bibinfo
  {title} {{Dynamical Heterogeneities in Glasses, Colloids, and Granular
  Media}}}}\ (\bibinfo  {publisher} {Oxford University Press},\ \bibinfo {year}
  {2011})\BibitemShut {NoStop}%
\bibitem [{\citenamefont {Scalliet}\ \emph {et~al.}(2022)\citenamefont
  {Scalliet}, \citenamefont {Guiselin},\ and\ \citenamefont
  {Berthier}}]{scalliet2022}%
  \BibitemOpen
  \bibfield  {author} {\bibinfo {author} {\bibfnamefont {C.}~\bibnamefont
  {Scalliet}}, \bibinfo {author} {\bibfnamefont {B.}~\bibnamefont {Guiselin}},\
  and\ \bibinfo {author} {\bibfnamefont {L.}~\bibnamefont {Berthier}},\
  }\bibfield  {title} {\bibinfo {title} {Thirty milliseconds in the life of a
  supercooled liquid},\ }\href {https://doi.org/10.1103/PhysRevX.12.041028}
  {\bibfield  {journal} {\bibinfo  {journal} {Physical Review X}\ }\textbf
  {\bibinfo {volume} {12}},\ \bibinfo {pages} {041028} (\bibinfo {year}
  {2022})}\BibitemShut {NoStop}%
\bibitem [{\citenamefont {Ediger}(2017)}]{ediger2017highly}%
  \BibitemOpen
  \bibfield  {author} {\bibinfo {author} {\bibfnamefont {M.~D.}\ \bibnamefont
  {Ediger}},\ }\bibfield  {title} {\bibinfo {title} {{Perspective: Highly
  stable vapor-deposited glasses}},\ }\href {https://doi.org/10.1063/1.5006265}
  {\bibfield  {journal} {\bibinfo  {journal} {The Journal of Chemical Physics}\
  }\textbf {\bibinfo {volume} {147}},\ \bibinfo {pages} {210901} (\bibinfo
  {year} {2017})}\BibitemShut {NoStop}%
\bibitem [{\citenamefont {Herrero}\ \emph
  {et~al.}(2023{\natexlab{a}})\citenamefont {Herrero}, \citenamefont {Ediger},\
  and\ \citenamefont {Berthier}}]{herrero2023front}%
  \BibitemOpen
  \bibfield  {author} {\bibinfo {author} {\bibfnamefont {C.}~\bibnamefont
  {Herrero}}, \bibinfo {author} {\bibfnamefont {M.~D.}\ \bibnamefont
  {Ediger}},\ and\ \bibinfo {author} {\bibfnamefont {L.}~\bibnamefont
  {Berthier}},\ }\bibfield  {title} {\bibinfo {title} {{Front propagation in
  ultrastable glasses is dynamically heterogeneous}},\ }\href
  {https://doi.org/10.1063/5.0168506} {\bibfield  {journal} {\bibinfo
  {journal} {The Journal of Chemical Physics}\ }\textbf {\bibinfo {volume}
  {159}},\ \bibinfo {pages} {114504} (\bibinfo {year}
  {2023}{\natexlab{a}})}\BibitemShut {NoStop}%
\bibitem [{\citenamefont {Léonard}\ and\ \citenamefont
  {Harrowell}(2010)}]{leonard2010}%
  \BibitemOpen
  \bibfield  {author} {\bibinfo {author} {\bibfnamefont {S.}~\bibnamefont
  {Léonard}}\ and\ \bibinfo {author} {\bibfnamefont {P.}~\bibnamefont
  {Harrowell}},\ }\bibfield  {title} {\bibinfo {title} {{Macroscopic
  facilitation of glassy relaxation kinetics: Ultrastable glass films with
  frontlike thermal response}},\ }\href {https://doi.org/10.1063/1.3511721}
  {\bibfield  {journal} {\bibinfo  {journal} {The Journal of Chemical Physics}\
  }\textbf {\bibinfo {volume} {133}},\ \bibinfo {pages} {244502} (\bibinfo
  {year} {2010})}\BibitemShut {NoStop}%
\bibitem [{\citenamefont {Sepúlveda}\ \emph {et~al.}(2013)\citenamefont
  {Sepúlveda}, \citenamefont {Swallen},\ and\ \citenamefont
  {Ediger}}]{sepulveda2013manipulating}%
  \BibitemOpen
  \bibfield  {author} {\bibinfo {author} {\bibfnamefont {A.}~\bibnamefont
  {Sepúlveda}}, \bibinfo {author} {\bibfnamefont {S.~F.}\ \bibnamefont
  {Swallen}},\ and\ \bibinfo {author} {\bibfnamefont {M.~D.}\ \bibnamefont
  {Ediger}},\ }\bibfield  {title} {\bibinfo {title} {{Manipulating the
  properties of stable organic glasses using kinetic facilitation}},\ }\href
  {https://doi.org/10.1063/1.4772594} {\bibfield  {journal} {\bibinfo
  {journal} {The Journal of Chemical Physics}\ }\textbf {\bibinfo {volume}
  {138}},\ \bibinfo {pages} {12A517} (\bibinfo {year} {2013})}\BibitemShut
  {NoStop}%
\bibitem [{\citenamefont {Herrero}\ \emph
  {et~al.}(2023{\natexlab{b}})\citenamefont {Herrero}, \citenamefont
  {Scalliet}, \citenamefont {Ediger},\ and\ \citenamefont
  {Berthier}}]{herrero2023two}%
  \BibitemOpen
  \bibfield  {author} {\bibinfo {author} {\bibfnamefont {C.}~\bibnamefont
  {Herrero}}, \bibinfo {author} {\bibfnamefont {C.}~\bibnamefont {Scalliet}},
  \bibinfo {author} {\bibfnamefont {M.~D.}\ \bibnamefont {Ediger}},\ and\
  \bibinfo {author} {\bibfnamefont {L.}~\bibnamefont {Berthier}},\ }\bibfield
  {title} {\bibinfo {title} {Two-step devitrification of ultrastable glasses},\
  }\href {https://doi.org/10.1073/pnas.2220824120} {\bibfield  {journal}
  {\bibinfo  {journal} {Proceedings of the National Academy of Sciences}\
  }\textbf {\bibinfo {volume} {120}},\ \bibinfo {pages} {e2220824120} (\bibinfo
  {year} {2023}{\natexlab{b}})}\BibitemShut {NoStop}%
\bibitem [{sup()}]{supplement}%
  \BibitemOpen
  \href@noop {} {\bibinfo {title} {See supplemental material at [url will be
  inserted by publisher] for additional descriptions of c.
  refs.~\cite{ozawa2019,larini2008,flenner2015,illing2017} are cited
  therein.}}\BibitemShut {Stop}%
\bibitem [{\citenamefont {Berthier}\ \emph {et~al.}(2017)\citenamefont
  {Berthier}, \citenamefont {Charbonneau}, \citenamefont {Coslovich},
  \citenamefont {Ninarello}, \citenamefont {Ozawa},\ and\ \citenamefont
  {Yaida}}]{berthier2017configurational}%
  \BibitemOpen
  \bibfield  {author} {\bibinfo {author} {\bibfnamefont {L.}~\bibnamefont
  {Berthier}}, \bibinfo {author} {\bibfnamefont {P.}~\bibnamefont
  {Charbonneau}}, \bibinfo {author} {\bibfnamefont {D.}~\bibnamefont
  {Coslovich}}, \bibinfo {author} {\bibfnamefont {A.}~\bibnamefont
  {Ninarello}}, \bibinfo {author} {\bibfnamefont {M.}~\bibnamefont {Ozawa}},\
  and\ \bibinfo {author} {\bibfnamefont {S.}~\bibnamefont {Yaida}},\ }\bibfield
   {title} {\bibinfo {title} {Configurational entropy measurements in extremely
  supercooled liquids that break the glass ceiling},\ }\href
  {https://doi.org/10.1073/pnas.1706860114} {\bibfield  {journal} {\bibinfo
  {journal} {Proceedings of the National Academy of Sciences}\ }\textbf
  {\bibinfo {volume} {114}},\ \bibinfo {pages} {11356} (\bibinfo {year}
  {2017})}\BibitemShut {NoStop}%
\bibitem [{\citenamefont {Berthier}\ \emph
  {et~al.}(2019{\natexlab{a}})\citenamefont {Berthier}, \citenamefont
  {Charbonneau}, \citenamefont {Ninarello}, \citenamefont {Ozawa},\ and\
  \citenamefont {Yaida}}]{berthier2019zero}%
  \BibitemOpen
  \bibfield  {author} {\bibinfo {author} {\bibfnamefont {L.}~\bibnamefont
  {Berthier}}, \bibinfo {author} {\bibfnamefont {P.}~\bibnamefont
  {Charbonneau}}, \bibinfo {author} {\bibfnamefont {A.}~\bibnamefont
  {Ninarello}}, \bibinfo {author} {\bibfnamefont {M.}~\bibnamefont {Ozawa}},\
  and\ \bibinfo {author} {\bibfnamefont {S.}~\bibnamefont {Yaida}},\ }\bibfield
   {title} {\bibinfo {title} {Zero-temperature glass transition in two
  dimensions},\ }\href {https://www.nature.com/articles/s41467-019-09512-3}
  {\bibfield  {journal} {\bibinfo  {journal} {Nature communications}\ }\textbf
  {\bibinfo {volume} {10}},\ \bibinfo {pages} {1508} (\bibinfo {year}
  {2019}{\natexlab{a}})}\BibitemShut {NoStop}%
\bibitem [{\citenamefont {Ninarello}\ \emph {et~al.}(2017)\citenamefont
  {Ninarello}, \citenamefont {Berthier},\ and\ \citenamefont
  {Coslovich}}]{ninarello2017}%
  \BibitemOpen
  \bibfield  {author} {\bibinfo {author} {\bibfnamefont {A.}~\bibnamefont
  {Ninarello}}, \bibinfo {author} {\bibfnamefont {L.}~\bibnamefont
  {Berthier}},\ and\ \bibinfo {author} {\bibfnamefont {D.}~\bibnamefont
  {Coslovich}},\ }\bibfield  {title} {\bibinfo {title} {Models and algorithms
  for the next generation of glass transition studies},\ }\href
  {https://doi.org/10.1103/PhysRevX.7.021039} {\bibfield  {journal} {\bibinfo
  {journal} {Phys. Rev. X}\ }\textbf {\bibinfo {volume} {7}},\ \bibinfo {pages}
  {021039} (\bibinfo {year} {2017})}\BibitemShut {NoStop}%
\bibitem [{\citenamefont {Berthier}\ \emph
  {et~al.}(2019{\natexlab{b}})\citenamefont {Berthier}, \citenamefont
  {Flenner}, \citenamefont {Fullerton}, \citenamefont {Scalliet},\ and\
  \citenamefont {Singh}}]{berthier2019efficient}%
  \BibitemOpen
  \bibfield  {author} {\bibinfo {author} {\bibfnamefont {L.}~\bibnamefont
  {Berthier}}, \bibinfo {author} {\bibfnamefont {E.}~\bibnamefont {Flenner}},
  \bibinfo {author} {\bibfnamefont {C.~J.}\ \bibnamefont {Fullerton}}, \bibinfo
  {author} {\bibfnamefont {C.}~\bibnamefont {Scalliet}},\ and\ \bibinfo
  {author} {\bibfnamefont {M.}~\bibnamefont {Singh}},\ }\bibfield  {title}
  {\bibinfo {title} {Efficient swap algorithms for molecular dynamics
  simulations of equilibrium supercooled liquids},\ }\href
  {https://iopscience.iop.org/article/10.1088/1742-5468/ab1910/meta?casa_token=5w8kUdks0hIAAAAA:ivu6CREeHaHpNQBYBqGaX_JoWzFHxFuKUwqXpK7eZAJnqM2W3QPIPS9ZXKM7mC4yWBlmRPm43wb9}
  {\bibfield  {journal} {\bibinfo  {journal} {Journal of Statistical Mechanics:
  Theory and Experiment}\ }\textbf {\bibinfo {volume} {2019}},\ \bibinfo
  {pages} {064004} (\bibinfo {year} {2019}{\natexlab{b}})}\BibitemShut
  {NoStop}%
\bibitem [{\citenamefont {Guiselin}\ \emph {et~al.}(2022)\citenamefont
  {Guiselin}, \citenamefont {Scalliet},\ and\ \citenamefont
  {Berthier}}]{guiselin2022}%
  \BibitemOpen
  \bibfield  {author} {\bibinfo {author} {\bibfnamefont {B.}~\bibnamefont
  {Guiselin}}, \bibinfo {author} {\bibfnamefont {C.}~\bibnamefont {Scalliet}},\
  and\ \bibinfo {author} {\bibfnamefont {L.}~\bibnamefont {Berthier}},\
  }\bibfield  {title} {\bibinfo {title} {Microscopic origin of excess wings in
  relaxation spectra of supercooled liquids},\ }\href
  {https://idp.nature.com/authorize/casa?redirect_uri=https://www.nature.com/articles/s41567-022-01508-z&casa_token=qtMCFKStZdwAAAAA:oGZaLDEEmmoFSbXHXndTmUeuZKZZnTF9U_i0Kz-31z3Lk3JbKJ1Hl8-2WjODKswbmV9JeKUGi9IhW7ob}
  {\bibfield  {journal} {\bibinfo  {journal} {Nature Physics}\ }\textbf
  {\bibinfo {volume} {18}},\ \bibinfo {pages} {468} (\bibinfo {year}
  {2022})}\BibitemShut {NoStop}%
\bibitem [{\citenamefont {Kob}\ \emph {et~al.}(2012)\citenamefont {Kob},
  \citenamefont {Rold{\'a}n-Vargas},\ and\ \citenamefont {Berthier}}]{kob2012}%
  \BibitemOpen
  \bibfield  {author} {\bibinfo {author} {\bibfnamefont {W.}~\bibnamefont
  {Kob}}, \bibinfo {author} {\bibfnamefont {S.}~\bibnamefont
  {Rold{\'a}n-Vargas}},\ and\ \bibinfo {author} {\bibfnamefont
  {L.}~\bibnamefont {Berthier}},\ }\bibfield  {title} {\bibinfo {title}
  {Non-monotonic temperature evolution of dynamic correlations in glass-forming
  liquids},\ }\href
  {https://idp.nature.com/authorize/casa?redirect_uri=https://www.nature.com/articles/nphys2133&casa_token=xRJBvx96iZUAAAAA:VS8KKc4XDPjPV1WpU3PHaXb5i7S9VJGwMtmcb7dVZvrSkguFNG9D0mJxCtqzCCwW1CRlaf1SH9RKz6Um}
  {\bibfield  {journal} {\bibinfo  {journal} {Nature Physics}\ }\textbf
  {\bibinfo {volume} {8}},\ \bibinfo {pages} {164} (\bibinfo {year}
  {2012})}\BibitemShut {NoStop}%
\bibitem [{\citenamefont {Berthier}\ and\ \citenamefont
  {Bouchaud}(2002)}]{berthier2002geometrical}%
  \BibitemOpen
  \bibfield  {author} {\bibinfo {author} {\bibfnamefont {L.}~\bibnamefont
  {Berthier}}\ and\ \bibinfo {author} {\bibfnamefont {J.-P.}\ \bibnamefont
  {Bouchaud}},\ }\bibfield  {title} {\bibinfo {title} {Geometrical aspects of
  aging and rejuvenation in the ising spin glass: A numerical study},\ }\href
  {https://doi.org/10.1103/PhysRevB.66.054404} {\bibfield  {journal} {\bibinfo
  {journal} {Phys. Rev. B}\ }\textbf {\bibinfo {volume} {66}},\ \bibinfo
  {pages} {054404} (\bibinfo {year} {2002})}\BibitemShut {NoStop}%
\bibitem [{\citenamefont {Dalle-Ferrier}\ \emph {et~al.}(2007)\citenamefont
  {Dalle-Ferrier}, \citenamefont {Thibierge}, \citenamefont {Alba-Simionesco},
  \citenamefont {Berthier}, \citenamefont {Biroli}, \citenamefont {Bouchaud},
  \citenamefont {Ladieu}, \citenamefont {L'H\^ote},\ and\ \citenamefont
  {Tarjus}}]{dalle2007spatial}%
  \BibitemOpen
  \bibfield  {author} {\bibinfo {author} {\bibfnamefont {C.}~\bibnamefont
  {Dalle-Ferrier}}, \bibinfo {author} {\bibfnamefont {C.}~\bibnamefont
  {Thibierge}}, \bibinfo {author} {\bibfnamefont {C.}~\bibnamefont
  {Alba-Simionesco}}, \bibinfo {author} {\bibfnamefont {L.}~\bibnamefont
  {Berthier}}, \bibinfo {author} {\bibfnamefont {G.}~\bibnamefont {Biroli}},
  \bibinfo {author} {\bibfnamefont {J.-P.}\ \bibnamefont {Bouchaud}}, \bibinfo
  {author} {\bibfnamefont {F.}~\bibnamefont {Ladieu}}, \bibinfo {author}
  {\bibfnamefont {D.}~\bibnamefont {L'H\^ote}},\ and\ \bibinfo {author}
  {\bibfnamefont {G.}~\bibnamefont {Tarjus}},\ }\bibfield  {title} {\bibinfo
  {title} {Spatial correlations in the dynamics of glassforming liquids:
  Experimental determination of their temperature dependence},\ }\href
  {https://doi.org/10.1103/PhysRevE.76.041510} {\bibfield  {journal} {\bibinfo
  {journal} {Phys. Rev. E}\ }\textbf {\bibinfo {volume} {76}},\ \bibinfo
  {pages} {041510} (\bibinfo {year} {2007})}\BibitemShut {NoStop}%
\bibitem [{\citenamefont {Berthier}\ and\ \citenamefont
  {Ediger}(2020)}]{berthier2020}%
  \BibitemOpen
  \bibfield  {author} {\bibinfo {author} {\bibfnamefont {L.}~\bibnamefont
  {Berthier}}\ and\ \bibinfo {author} {\bibfnamefont {M.~D.}\ \bibnamefont
  {Ediger}},\ }\bibfield  {title} {\bibinfo {title} {How to “measure” a
  structural relaxation time that is too long to be measured?},\ }\href
  {https://pubs.aip.org/aip/jcp/article-abstract/153/4/044501/1065001}
  {\bibfield  {journal} {\bibinfo  {journal} {The Journal of Chemical Physics}\
  }\textbf {\bibinfo {volume} {153}} (\bibinfo {year} {2020})}\BibitemShut
  {NoStop}%
\bibitem [{\citenamefont {Ortlieb}\ \emph {et~al.}(2023)\citenamefont
  {Ortlieb}, \citenamefont {Ingebrigtsen}, \citenamefont {Hallett},
  \citenamefont {Turci},\ and\ \citenamefont {Royall}}]{ortlieb2023}%
  \BibitemOpen
  \bibfield  {author} {\bibinfo {author} {\bibfnamefont {L.}~\bibnamefont
  {Ortlieb}}, \bibinfo {author} {\bibfnamefont {T.~S.}\ \bibnamefont
  {Ingebrigtsen}}, \bibinfo {author} {\bibfnamefont {J.~E.}\ \bibnamefont
  {Hallett}}, \bibinfo {author} {\bibfnamefont {F.}~\bibnamefont {Turci}},\
  and\ \bibinfo {author} {\bibfnamefont {C.~P.}\ \bibnamefont {Royall}},\
  }\bibfield  {title} {\bibinfo {title} {Probing excitations and cooperatively
  rearranging regions in deeply supercooled liquids},\ }\href
  {https://doi.org/10.1038/s41467-023-37793-2} {\bibfield  {journal} {\bibinfo
  {journal} {Nature Communications}\ }\textbf {\bibinfo {volume} {14}},\
  \bibinfo {pages} {2621} (\bibinfo {year} {2023})}\BibitemShut {NoStop}%
\bibitem [{\citenamefont {B\"assler}(1987)}]{bassler1987viscous}%
  \BibitemOpen
  \bibfield  {author} {\bibinfo {author} {\bibfnamefont {H.}~\bibnamefont
  {B\"assler}},\ }\bibfield  {title} {\bibinfo {title} {Viscous flow in
  supercooled liquids analyzed in terms of transport theory for random media
  with energetic disorder},\ }\href
  {https://doi.org/10.1103/PhysRevLett.58.767} {\bibfield  {journal} {\bibinfo
  {journal} {Phys. Rev. Lett.}\ }\textbf {\bibinfo {volume} {58}},\ \bibinfo
  {pages} {767} (\bibinfo {year} {1987})}\BibitemShut {NoStop}%
\bibitem [{\citenamefont {Sussman}\ \emph {et~al.}(2017)\citenamefont
  {Sussman}, \citenamefont {Schoenholz}, \citenamefont {Cubuk},\ and\
  \citenamefont {Liu}}]{sussman2017}%
  \BibitemOpen
  \bibfield  {author} {\bibinfo {author} {\bibfnamefont {D.~M.}\ \bibnamefont
  {Sussman}}, \bibinfo {author} {\bibfnamefont {S.~S.}\ \bibnamefont
  {Schoenholz}}, \bibinfo {author} {\bibfnamefont {E.~D.}\ \bibnamefont
  {Cubuk}},\ and\ \bibinfo {author} {\bibfnamefont {A.~J.}\ \bibnamefont
  {Liu}},\ }\bibfield  {title} {\bibinfo {title} {Disconnecting structure and
  dynamics in glassy thin films},\ }\href
  {https://doi.org/10.1073/pnas.1703927114} {\bibfield  {journal} {\bibinfo
  {journal} {Proceedings of the National Academy of Sciences}\ }\textbf
  {\bibinfo {volume} {114}},\ \bibinfo {pages} {10601} (\bibinfo {year}
  {2017})}\BibitemShut {NoStop}%
\bibitem [{\citenamefont {Das}\ and\ \citenamefont
  {Sastry}(2022)}]{DAS2022100098}%
  \BibitemOpen
  \bibfield  {author} {\bibinfo {author} {\bibfnamefont {P.}~\bibnamefont
  {Das}}\ and\ \bibinfo {author} {\bibfnamefont {S.}~\bibnamefont {Sastry}},\
  }\bibfield  {title} {\bibinfo {title} {Crossover in dynamics in the
  kob-andersen binary mixture glass-forming liquid},\ }\href
  {https://doi.org/https://doi.org/10.1016/j.nocx.2022.100098} {\bibfield
  {journal} {\bibinfo  {journal} {Journal of Non-Crystalline Solids: X}\
  }\textbf {\bibinfo {volume} {14}},\ \bibinfo {pages} {100098} (\bibinfo
  {year} {2022})}\BibitemShut {NoStop}%
\bibitem [{\citenamefont {Jung}\ \emph {et~al.}(2023)\citenamefont {Jung},
  \citenamefont {Biroli},\ and\ \citenamefont {Berthier}}]{jung2023predicting}%
  \BibitemOpen
  \bibfield  {author} {\bibinfo {author} {\bibfnamefont {G.}~\bibnamefont
  {Jung}}, \bibinfo {author} {\bibfnamefont {G.}~\bibnamefont {Biroli}},\ and\
  \bibinfo {author} {\bibfnamefont {L.}~\bibnamefont {Berthier}},\ }\bibfield
  {title} {\bibinfo {title} {Predicting dynamic heterogeneity in glass-forming
  liquids by physics-inspired machine learning},\ }\href
  {https://doi.org/10.1103/PhysRevLett.130.238202} {\bibfield  {journal}
  {\bibinfo  {journal} {Phys. Rev. Lett.}\ }\textbf {\bibinfo {volume} {130}},\
  \bibinfo {pages} {238202} (\bibinfo {year} {2023})}\BibitemShut {NoStop}%
\bibitem [{\citenamefont {Kirkpatrick}\ \emph {et~al.}(1989)\citenamefont
  {Kirkpatrick}, \citenamefont {Thirumalai},\ and\ \citenamefont
  {Wolynes}}]{kirkpatrick1989}%
  \BibitemOpen
  \bibfield  {author} {\bibinfo {author} {\bibfnamefont {T.~R.}\ \bibnamefont
  {Kirkpatrick}}, \bibinfo {author} {\bibfnamefont {D.}~\bibnamefont
  {Thirumalai}},\ and\ \bibinfo {author} {\bibfnamefont {P.~G.}\ \bibnamefont
  {Wolynes}},\ }\bibfield  {title} {\bibinfo {title} {Scaling concepts for the
  dynamics of viscous liquids near an ideal glassy state},\ }\href
  {https://doi.org/10.1103/PhysRevA.40.1045} {\bibfield  {journal} {\bibinfo
  {journal} {Phys. Rev. A}\ }\textbf {\bibinfo {volume} {40}},\ \bibinfo
  {pages} {1045} (\bibinfo {year} {1989})}\BibitemShut {NoStop}%
\bibitem [{\citenamefont {Bouchaud}\ and\ \citenamefont
  {Biroli}(2004)}]{bouchaud2004}%
  \BibitemOpen
  \bibfield  {author} {\bibinfo {author} {\bibfnamefont {J.-P.}\ \bibnamefont
  {Bouchaud}}\ and\ \bibinfo {author} {\bibfnamefont {G.}~\bibnamefont
  {Biroli}},\ }\bibfield  {title} {\bibinfo {title} {{On the
  Adam-Gibbs-Kirkpatrick-Thirumalai-Wolynes scenario for the viscosity increase
  in glasses}},\ }\href {https://doi.org/10.1063/1.1796231} {\bibfield
  {journal} {\bibinfo  {journal} {The Journal of Chemical Physics}\ }\textbf
  {\bibinfo {volume} {121}},\ \bibinfo {pages} {7347} (\bibinfo {year}
  {2004})}\BibitemShut {NoStop}%
\bibitem [{\citenamefont {Berthier}(2021)}]{berthier2021}%
  \BibitemOpen
  \bibfield  {author} {\bibinfo {author} {\bibfnamefont {L.}~\bibnamefont
  {Berthier}},\ }\bibfield  {title} {\bibinfo {title} {Self-induced
  heterogeneity in deeply supercooled liquids},\ }\href
  {https://journals.aps.org/prl/abstract/10.1103/PhysRevLett.127.088002}
  {\bibfield  {journal} {\bibinfo  {journal} {Physical Review Letters}\
  }\textbf {\bibinfo {volume} {127}},\ \bibinfo {pages} {088002} (\bibinfo
  {year} {2021})}\BibitemShut {NoStop}%
\bibitem [{\citenamefont {Ozawa}\ \emph {et~al.}(2019)\citenamefont {Ozawa},
  \citenamefont {Scalliet}, \citenamefont {Ninarello},\ and\ \citenamefont
  {Berthier}}]{ozawa2019}%
  \BibitemOpen
  \bibfield  {author} {\bibinfo {author} {\bibfnamefont {M.}~\bibnamefont
  {Ozawa}}, \bibinfo {author} {\bibfnamefont {C.}~\bibnamefont {Scalliet}},
  \bibinfo {author} {\bibfnamefont {A.}~\bibnamefont {Ninarello}},\ and\
  \bibinfo {author} {\bibfnamefont {L.}~\bibnamefont {Berthier}},\ }\bibfield
  {title} {\bibinfo {title} {Does the adam-gibbs relation hold in simulated
  supercooled liquids?},\ }\bibfield  {journal} {\bibinfo  {journal} {The
  Journal of chemical physics}\ }\textbf {\bibinfo {volume} {151}},\ \href
  {https://doi.org/https://doi.org/10.1063/1.5113477}
  {https://doi.org/10.1063/1.5113477} (\bibinfo {year} {2019})\BibitemShut
  {NoStop}%
\bibitem [{\citenamefont {Larini}\ \emph {et~al.}(2008)\citenamefont {Larini},
  \citenamefont {Ottochian}, \citenamefont {De~Michele},\ and\ \citenamefont
  {Leporini}}]{larini2008}%
  \BibitemOpen
  \bibfield  {author} {\bibinfo {author} {\bibfnamefont {L.}~\bibnamefont
  {Larini}}, \bibinfo {author} {\bibfnamefont {A.}~\bibnamefont {Ottochian}},
  \bibinfo {author} {\bibfnamefont {C.}~\bibnamefont {De~Michele}},\ and\
  \bibinfo {author} {\bibfnamefont {D.}~\bibnamefont {Leporini}},\ }\bibfield
  {title} {\bibinfo {title} {Universal scaling between structural relaxation
  and vibrational dynamics in glass-forming liquids and polymers},\ }\href
  {https://www.nature.com/articles/nphys788} {\bibfield  {journal} {\bibinfo
  {journal} {Nature Physics}\ }\textbf {\bibinfo {volume} {4}},\ \bibinfo
  {pages} {42} (\bibinfo {year} {2008})}\BibitemShut {NoStop}%
\bibitem [{\citenamefont {Flenner}\ and\ \citenamefont
  {Szamel}(2015)}]{flenner2015}%
  \BibitemOpen
  \bibfield  {author} {\bibinfo {author} {\bibfnamefont {E.}~\bibnamefont
  {Flenner}}\ and\ \bibinfo {author} {\bibfnamefont {G.}~\bibnamefont
  {Szamel}},\ }\bibfield  {title} {\bibinfo {title} {Fundamental differences
  between glassy dynamics in two and three dimensions},\ }\href
  {https://www.nature.com/articles/ncomms8392} {\bibfield  {journal} {\bibinfo
  {journal} {Nature communications}\ }\textbf {\bibinfo {volume} {6}},\
  \bibinfo {pages} {7392} (\bibinfo {year} {2015})}\BibitemShut {NoStop}%
\bibitem [{\citenamefont {Illing}\ \emph {et~al.}(2017)\citenamefont {Illing},
  \citenamefont {Fritschi}, \citenamefont {Kaiser}, \citenamefont {Klix},
  \citenamefont {Maret},\ and\ \citenamefont {Keim}}]{illing2017}%
  \BibitemOpen
  \bibfield  {author} {\bibinfo {author} {\bibfnamefont {B.}~\bibnamefont
  {Illing}}, \bibinfo {author} {\bibfnamefont {S.}~\bibnamefont {Fritschi}},
  \bibinfo {author} {\bibfnamefont {H.}~\bibnamefont {Kaiser}}, \bibinfo
  {author} {\bibfnamefont {C.~L.}\ \bibnamefont {Klix}}, \bibinfo {author}
  {\bibfnamefont {G.}~\bibnamefont {Maret}},\ and\ \bibinfo {author}
  {\bibfnamefont {P.}~\bibnamefont {Keim}},\ }\bibfield  {title} {\bibinfo
  {title} {Mermin--wagner fluctuations in 2d amorphous solids},\ }\href
  {https://www.pnas.org/doi/abs/10.1073/pnas.1612964114} {\bibfield  {journal}
  {\bibinfo  {journal} {Proceedings of the National Academy of Sciences}\
  }\textbf {\bibinfo {volume} {114}},\ \bibinfo {pages} {1856} (\bibinfo {year}
  {2017})}\BibitemShut {NoStop}%
\end{thebibliography}
